\documentclass[journal]{IEEEtran}

\ifCLASSOPTIONcompsoc
  \usepackage[nocompress]{cite}
\else
  \usepackage{cite}
\fi
\usepackage[ruled,vlined]{algorithm2e}
\usepackage{pifont}
\usepackage{cite}
\usepackage{amsmath,amssymb,amsfonts}
\usepackage{stfloats}
\usepackage{graphicx}
\usepackage{textcomp}
\usepackage{xcolor}
\usepackage{mathrsfs}
\usepackage{amsmath}
\usepackage{mathfixs}
\usepackage{subfigure}
\usepackage{float}
\usepackage{amsthm}
\usepackage{balance}

\usepackage[T1]{fontenc}
\usepackage{ragged2e}
\usepackage{bm}
\usepackage{setspace}

\newcommand{\name}{SUAD}
\newcommand{\revr}{\textcolor[rgb]{0,0,0}}
\newcommand{\revc}{\textcolor[rgb]{0,0,0}}

\begin{document}
\title{SUAD: Solid-Channel Ultrasound Injection Attack and Defense to Voice Assistants}

\author{Chao Liu,~\IEEEmembership{Senior~Member,~IEEE},
        Zhezheng~Zhu,
        Hao Chen,
        Kaiwen Guo,
        
        Penghao Wang,
        and~Xiang-Yang Li,~\IEEEmembership{Fellow,~IEEE}
\IEEEcompsocitemizethanks{%
\IEEEcompsocthanksitem
This research was funded by the National Natural Science Foundation of China grant No. 62272427, Shandong Provincial Natural Science Foundation No. ZR2025QB43, ZR2025QC1536, Hainan Provincial Natural Science Foundation No. 626YXQN0945. \textit{(Corresponding authors: Penghao Wang.)}
\IEEEcompsocthanksitem
Chao Liu is with the Department of Computer Science and Technology, Ocean University of China, Qingdao 266100, China, and with the Sanya Ocean Institute, Ocean University of China, Sanya 572000, China. (e-mail: liuchao@ouc.edu.cn)
\IEEEcompsocthanksitem
Zhezheng Zhu, Hao Chen and Kaiwen Guo are with the Department of Computer Science and Technology, Ocean University of China, Qingdao 266100, China (e-mail: zhuzhezheng@stu.ouc.edu.cn;  chenhao1339@stu.ouc.edu.cn; kevinguo@ouc.edu.cn).
\IEEEcompsocthanksitem
Penghao Wang is currently a Research Fellow with the School of Computer Science and Engineering, Nanyang Technological University, Singapore 639798.(e-mail:wangpenghao@stu.ouc.edu.cn) 
\IEEEcompsocthanksitem
Xiang-Yang Li is with the School of Computer Science and Technology, University of Science and Technology of China, Hefei 230026, China (e-mail: xiangyangli@ustc.edu.cn).}
}

\markboth{IEEE TRANSACTIONS ON MOBILE COMPUTING,~Vol.~XX, No.~X, XXXX~202X}%
{Shell \MakeLowercase{\textit{et al.}}: Bare Demo of IEEEtran.cls for IEEE Journals}

\maketitle
\begin{abstract}
As a versatile AI application, voice assistants (VAs) have become increasingly popular, but are vulnerable to security threats. Attackers have proposed various inaudible attacks, but are limited by distance, or Line of Sight (LoS). Therefore, we propose \name~Attack, a long-range, cross-barrier, and interference-free inaudible voice attack via solid channels. We begin by thoroughly analyzing the dispersion effect in solid channels, revealing its unique impact on signal propagation. To avoid distortions in voice commands, we design a low-overhead modular command generation model that parameterizes attack distance, victim audio, and medium dispersion features for real-time adaptation to variations in the solid-channel state. 
Additionally, we propose SUAD Defense, a universal defense that uses ultrasonic perturbation signals to block inaudible voice attacks (IVAs) without impacting normal speech. Since the attack can occur at arbitrary frequencies and times, we propose a training method that randomizes both time and frequency to generate perturbation signals that break ultrasonic commands. Notably, the perturbation signal is modulated to an inaudible frequency without affecting the functionality of voice commands for VAs.
Experiments on six smartphones have shown that SUAD Attack achieves activation success rates above 89.8\% and SUAD Defense blocks IVAs with success rates exceeding 98\%.
\end{abstract}

\begin{IEEEkeywords}
Microphone Nonlinearity, Acoustic Dispersion, Inaudible Voice Attack, Universal Adversarial Perturbation
\end{IEEEkeywords}

\IEEEpeerreviewmaketitle

\section{Introduction}
\IEEEPARstart{R}{apid} advances in Artificial Intelligence (AI) have contributed to the widespread application of Voice Assistants (VAs) \cite{VAs, AI_survey}. With system privileges granted by users, VAs, such as Siri and Bixby, boast an extensive array of functionalities, including making calls, controlling device settings, and retrieving information \cite{Siri}. However, the gradual expansion of functions brings not only convenience but also security risks.
They are vulnerable to attacks involving forged voice commands, which can trigger high-risk operations such as unauthorized retrieval and viewing of private information or unverified payments and transfers \cite{Voice_Payment}, posing significant risks to users' information and property security. Therefore, conducting a comprehensive security analysis of voice command attacks on VAs is critically important.

\begin{figure}[t]
    \centering
    \includegraphics[width=\linewidth]{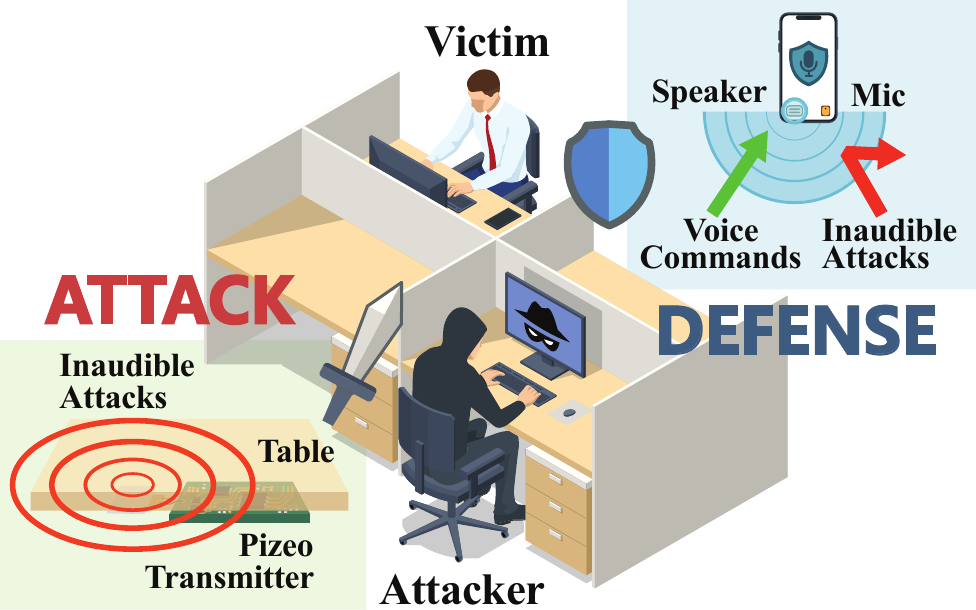}
    \caption{\revc{SUAD: (1) Attack: A piezo-transmitter covertly placed beneath a table emits inaudible signals to attack VAs. (2) Defense: Inaudible perturbation signals block such attacks without interfering with normal voice commands.}}
    \label{fig:senario}
\end{figure}

In recent years, attackers have proposed various methods \cite{Transduction_Shield, Inaudible_Attack2021, MagBackdoor, dai2023inducing,Ghost13, IEMI15, GhostTalk22, Light_Commands, zhang2017dolphinattack, roy2017backdoor, Lipread_2018, yan2020surfingattack, CapSpeaker_2021} to enable inaudible voice command injection. For example, Dai et al.~ \cite{dai2023inducing} generate magnetic-inductive voice signals by exploiting vulnerabilities in wireless charger systems or hardware, but the magnetic field restricts the range in 5~cm. Additionally, Light Commands \cite{Light_Commands} leverage laser beams pre-encoded with voice commands, which shine on microphones to produce spoofed electrical signals (i.e., audio). Despite its long attack range (110~m), high-cost devices and light-of-sight (LoS) greatly limit its application. 

Acknowledging these limitations, attackers have explored the feasibility of \textit{inaudible voice attacks} (IVAs) \cite{zhang2017dolphinattack, roy2017backdoor, Lipread_2018, yan2020surfingattack, CapSpeaker_2021}. DolphinAttack \cite{zhang2017dolphinattack} and BackDoor \cite{roy2017backdoor} exploit microphone nonlinearity to convert voice commands modulated at inaudible frequencies (\textgreater\,20\,kHz) into audible signals, enabling attacks at ranges of 1.75\,m and 3.5\,m, respectively. For longer range attacks (7.62\,m), LipRead \cite{Lipread_2018} utilizes different speakers in the array to play segments of the voice command spectrum, but sonic focusing requires LoS. Moreover, they share a common limitation: barriers can obstruct air-channel signal transmission. In fact, sound waves can propagate through any medium (e.g., solid) that supports vibrations. Recently, SurfingAttack\cite{yan2020surfingattack}, generates ultrasound on a solid surface, which propagates through the solid medium, to induce a nonlinear response in the microphone, injecting inaudible commands. Despite offering a clear analysis of solid-channel signal propagation, it neglects \textit{voice command distortion resulting from frequency-dependent dispersion}.

In this work, we aim to break this old notion by proposing a novel inaudible voice attack against solid-channel interference. As illustrated in Fig.~\ref{fig:senario}, a piezoelectric transmitter covertly affixed by the attacker under the table induces vibrations to emit ultrasonic attack signals encoded with malicious commands (e.g., 'upload album to Facebook'). Then, these signals are transmitted to the victim's device through the table, effectively bypassing LoS barriers. However, to effectively inject voice commands, the attack must overcome challenges inherent in solid-channel transmission. Specifically, due to the dispersion effect \cite{dispersion}, signals of different frequencies propagate at varying speeds in solid media, causing the original waveform features to change. On one hand, propagation over a distance introduces time delays that vary across different frequency components. On the other hand, signal waveforms gradually deform or broaden during propagation, resulting in significant distortion of the attack command.

\revr{To this end, we propose a novel solid-channel ultrasound attack and defense, i.e., SUAD. The SUAD Attack overcomes}
the aforementioned technical bottleneck and enables long-distance, cross-barrier, and interference-free inaudible attacks on VAs. We first employ a localization method to locate the victim device position by analyzing the time difference of arrival (TDoA) of solid-channel sound captured by a microphone array. Next, the estimated attack distance is used to compensate for propagation delays. Specifically, we use a speech cloning model with a fused multi-head architecture to generate attack commands capable of bypassing voiceprint authentication on VAs. These commands embed the victim's voiceprint features and inverse solid-channel interference, derived from the captured victim speech and the solid medium's physical characteristics, including propagation distance and material density, and varying board thicknesses. Ultimately, attack commands are modulated into the ultrasonic frequency band, generating inaudible signals that threaten VAs via solid-channel propagation.

With increasing awareness of IVA threats, various defenses \cite{MicGuard2024, DualGuard20, Cacher19, zhou2019hidden, VAuth17, guan2023trustworthy,EarArray,RobustDetection21,arrayID22,zhang2017dolphinattack, Lipread_2018, Watchdog2020 } have been developed against such attacks. For instance, DolphinAttack \cite{zhang2017dolphinattack} employs support vector machines (SVM) to classify and detect malicious audio, while MicGuard \cite{MicGuard2024} identifies anomalies in the spectral domain. Although these approaches can alert users, attacks may have been executed successfully before detection. Furthermore, systems like DualGuard \cite{DualGuard20} and Cacher \cite{Cacher19} may disable the microphone upon detecting an attack, potentially disrupting the normal operation of VAs. This poses a critical problem: \textit{How can we ensure the normal functionality of VAs while simultaneously enabling real-time defense against IVAs}?

\revc{As such, we propose SUAD Defense, which continuously emits ultrasonic perturbation signals via the smartphone’s speaker to actively defend against randomly launched, arbitrary-frequency IVAs. The short mic–speaker distance also allows its low-power operation. The perturbation is generated through a Universal Adversarial Perturbation (UAP) training method that randomizes both the temporal shift and frequency of attack signals during model training. As a result, the generated signal can effectively suppress attacks while allowing legitimate user voice commands to function correctly.}

\revc{Finally, we conducted extensive experiments to evaluate the effectiveness of SUAD Attack and Defense. SUAD Attack achieved median activation success rates of over 89.8\% across six smartphones. SUAD Defense has defense success rates exceeding 98\% against three types of attacks. In summary, the main contributions of this paper are as follows:}
\begin{itemize}
    \item \revc{We propose a novel \name~Attack, the first inaudible attack capable of adapting to solid-channel states. It overturns the old notion that solid-channel propagation is distortion-free,  extending the attack range.}
    \item \revc{A speech generation model with a fused multi-head architecture is proposed to embed the victim's voiceprint features and inverse solid-channel interference into voice commands, enabling successful activation on VAs.}
    \item \revc{We propose \name~Defense by designing a novel UAP training framework to generate universal perturbation signals. It defends against attacks launched randomly at any frequency, without affecting VAs.}
    \item Finally, extensive evaluations of \name~Attack and Defense demonstrate attack and defense success rates exceeding 89.8\% and 98\%, respectively.

\end{itemize}

\section{Background and Motivation}
\revc{In this section, we first present the threat model of VAs attacks. Next, we discuss the advantages and challenges of solid-channel attacks. Finally, we elaborate on the technical challenges of implementing defenses under nonlinear effects.}

\subsection{Threat Model} \label{Threat_Model}
We consider a common scenario where a victim, Bob, has an intelligent voice assistant installed on his smart devices. Due to frequent use, Bob often leaves the device active on the table, freeing his hands. This provides an opportunity for the attacker, Eve, to compromise the system and threaten information security. However, vigilant Bob may avoid suspicious objects within the LoS. In other words, Bob prefers familiar and private environments, such as a covered table, which obstructs voice attacks via air channels. It compels Eve to exploit other channels for launching attacks.

\begin{figure}[b]
    \setlength\abovecaptionskip{0pt}
    \setlength{\subfigcapskip}{0pt}
    \centering
    \includegraphics[width=\linewidth]{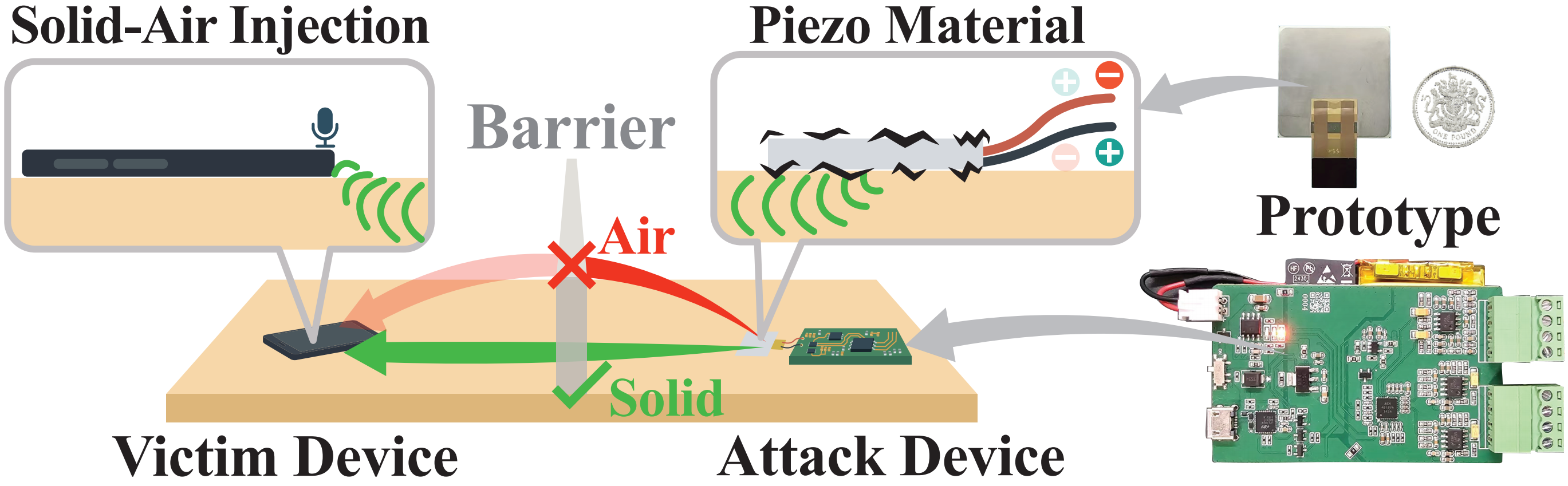}
    \vspace{1em}
    \caption{\revc{Comparison of air-channel and solid-channel voice injection.}}
    \label{fig: dispersion}
\end{figure}

In the aforementioned scenario, it is assumed that Eve can covertly place a small, autonomously operating attack device within a hidden physical space (e.g., under-table space), as illustrated in Fig.~\ref{fig:senario}, without alerting Bob. Thus, Eve can't remain on-site or nearby, avoiding any suspicious interaction. Moreover, Eve can generate arbitrary voice commands with voiceprint features by leveraging known voice samples of the victim. Fake commands in VAs attacks may include sensitive operations, such as 'turn on camera,' exploiting legitimate privileges to bypass protections and heighten security risks.

\begin{figure*}[t]
    \setlength\abovecaptionskip{0pt}
    \setlength{\subfigcapskip}{0pt}
    \centering
    \subfigure[Experimental setup for capturing audio from different channels]{    
        \centering
        \includegraphics[height =.145\linewidth]{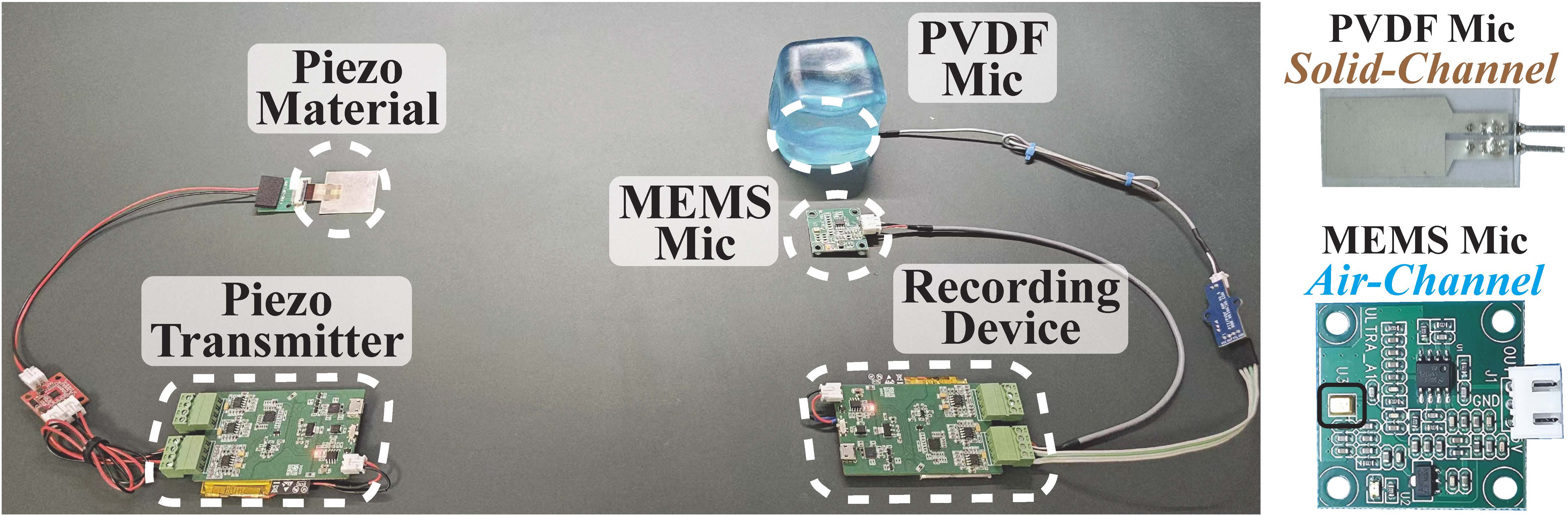}
        \label{fig:exsetup}
    }
    \subfigure[Raw audio]{    
        \centering
        \includegraphics[height =.145\linewidth]{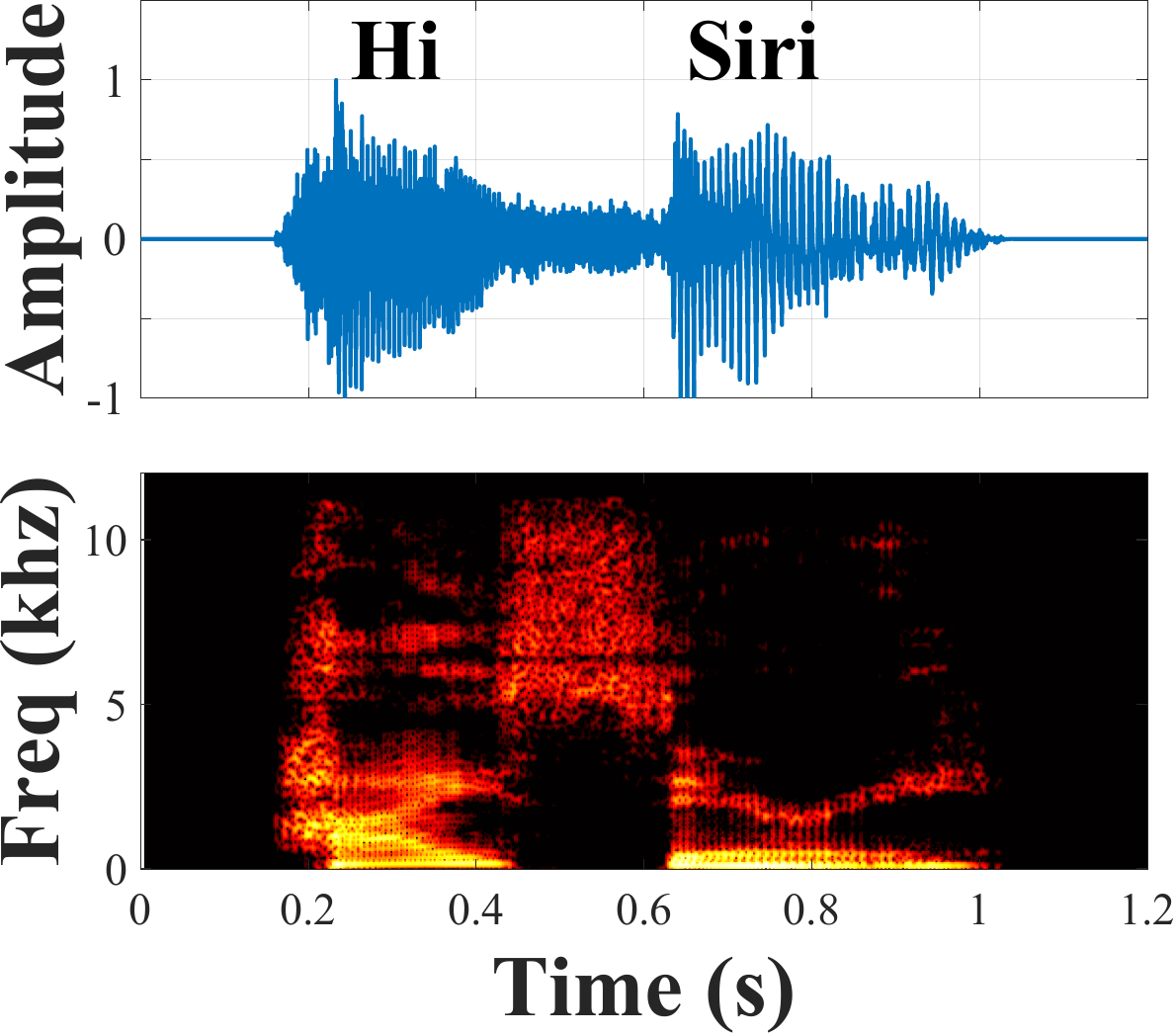}
        \label{fig:dispersion_voice1}
    }
    \subfigure[Air-channel]{
        \centering
        \includegraphics[height = .145\linewidth]{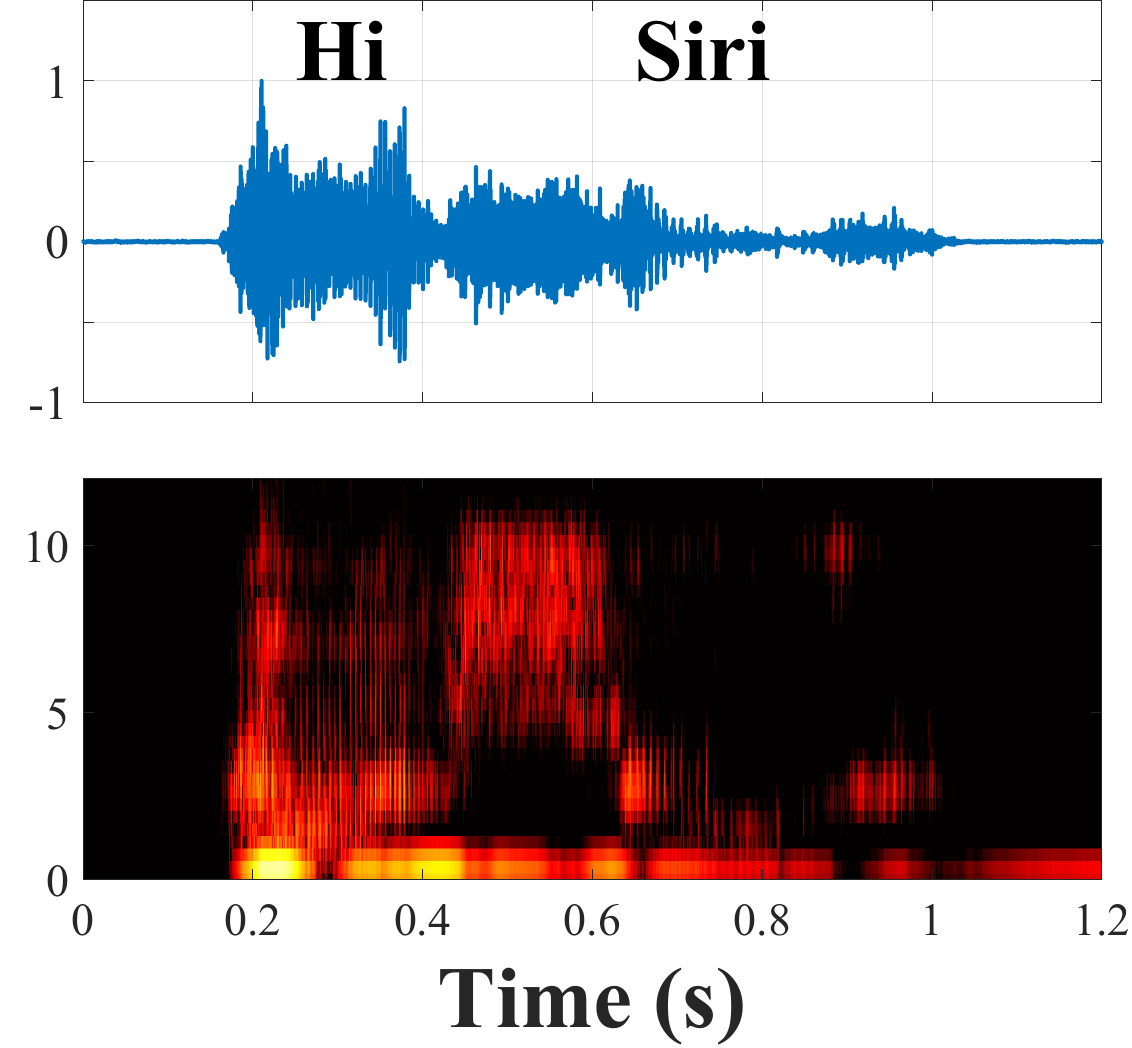}
        \label{fig:dispersion_voice2}
    }
    \subfigure[Solid-channel]{
        \centering
        \includegraphics[height = .145\linewidth]{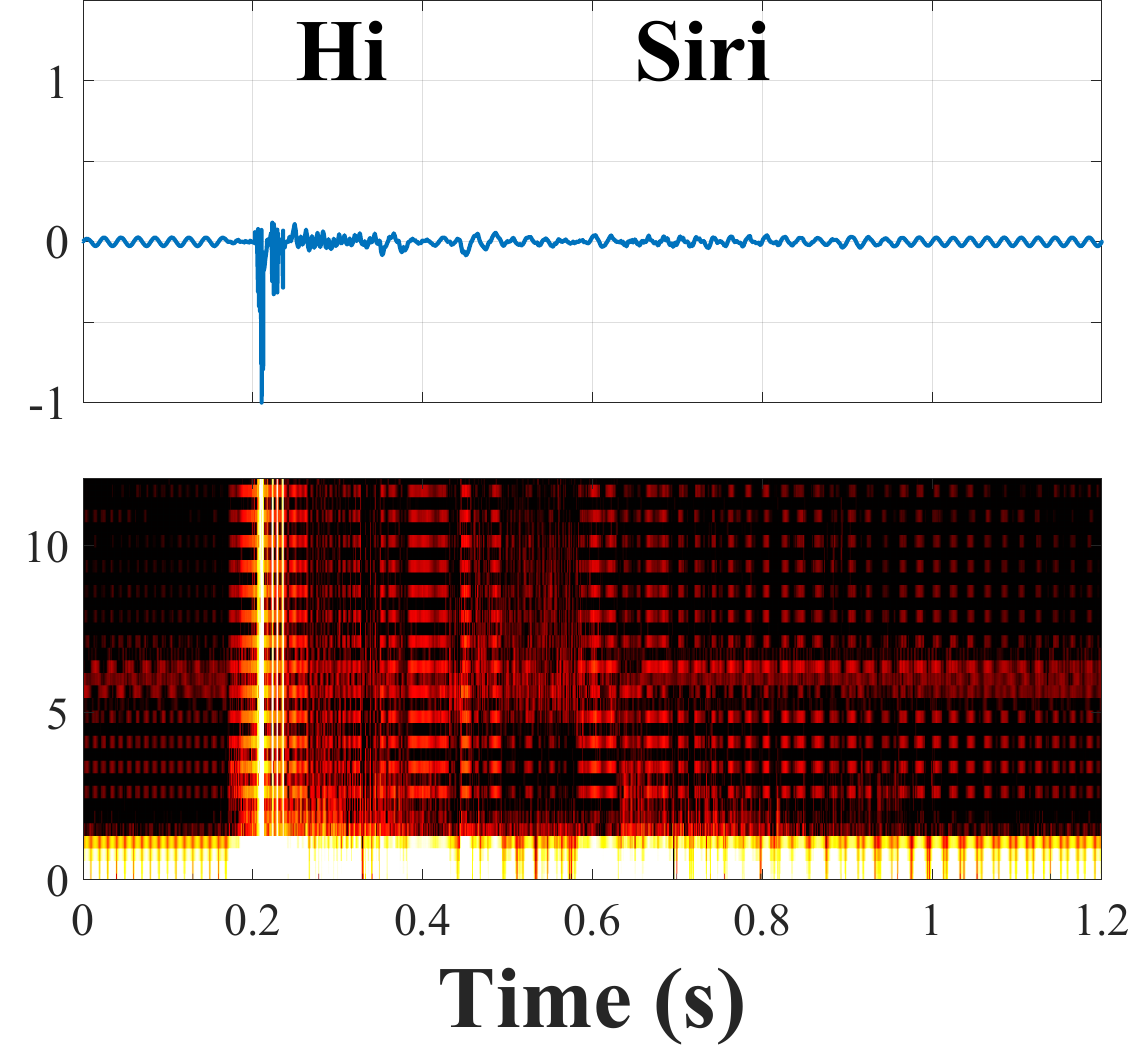}
        \label{fig:dispersion_voice3}
    }
    \caption{\revc{Comparison of audio through different-channel propagation}.}
    \label{fig:dispersion_voice}
    \vspace{-1.5em}
\end{figure*}

\subsection{\revc{Why Solid Instead of Air Channels?}}
\revc{Traditional attacks typically rely on ultrasonic propagation through the air. However, as illustrated in Fig.~\ref{fig: dispersion}, real-world scenarios may involve barriers obstructing the LoS path, making the solid channel (e.g., table) a viable alternative. Therefore, \name~Attack utilizes a piezoelectric element tightly coupled to the table underside as the attack transmitter, with its compact design enabling covert deployment. Due to the piezoelectric effect, applying an alternating voltage across the piezoelectric material makes slight deformations in its internal lattice structure, generating vibrations (i.e., sound). As such, the piezoelectric element can emit ultrasound carrying the attack command, which propagates through solid channels and then into the air, reaching the target device’s microphone to enable voice command injection, as shown in Fig.~\ref{fig: dispersion}.}

\revc{Compared to air channels, solid channels exhibit a distinct physical phenomenon known as the dispersion effect. This effect causes sound waves of different frequencies to propagate at varying speeds within solids, leading to gradual waveform deformation or broadening during transmission. To investigate the impact of dispersion on signal propagation, we use the piezoelectric material in the transmitter shown in Fig.~\ref{fig:exsetup} as the sound source to vibrate the table and emit the audio command 'Hi Siri'. The signal propagates through the table and is recorded by a MEMS microphone and a PVDF microphone, capturing the signals transmitted via the air and solid channels, respectively.
As shown in Fig.~\ref{fig:dispersion_voice3}, audio transmitted through the solid channel undergoes significant dispersion in both time and frequency domains. Even after re-transmission from the solid channel to air, the waveform remains notably altered, and the spectral components exhibit varying degrees of distortion, as shown in Fig.~\ref{fig:dispersion_voice2}. The effect on wave speed can be approximated by the Lamb wave dispersion equation, i.e.,} 
\begin{equation}\label{eq1}
    v(f) = \sqrt[4]{\frac{E h f^2}{12 \rho \left(1 - v_p^2\right)}},
\end{equation}
\revc{where $E$, $\rho$, $h$, and $f$ denote the Young’s modulus, material density, plate thickness, and frequency, respectively, and $v_p$ represents the Poisson’s ratio. } However, complex and various parameters make modeling $v(f)$ across all realistic scenarios impractical\cite{yan2020surfingattack}.

\revc{Notably, successful voice injection attacks via solid channels must overcome two key challenges: (1) variations in medium materials and (2) time delays caused by propagation distances.
Therefore, \name~Attack employs a two-stage synergetic scheme to embed inverse solid-channel propagation features into generated voice injection commands, mitigating signal distortion of voice commands. The specific design and implementation details will be developed in Section~\ref{SUAD_Attack}.}

\begin{figure}[b]
    \centering
    \subfigure[Raw audio]{    
        \centering
        \includegraphics[height =.27\linewidth]{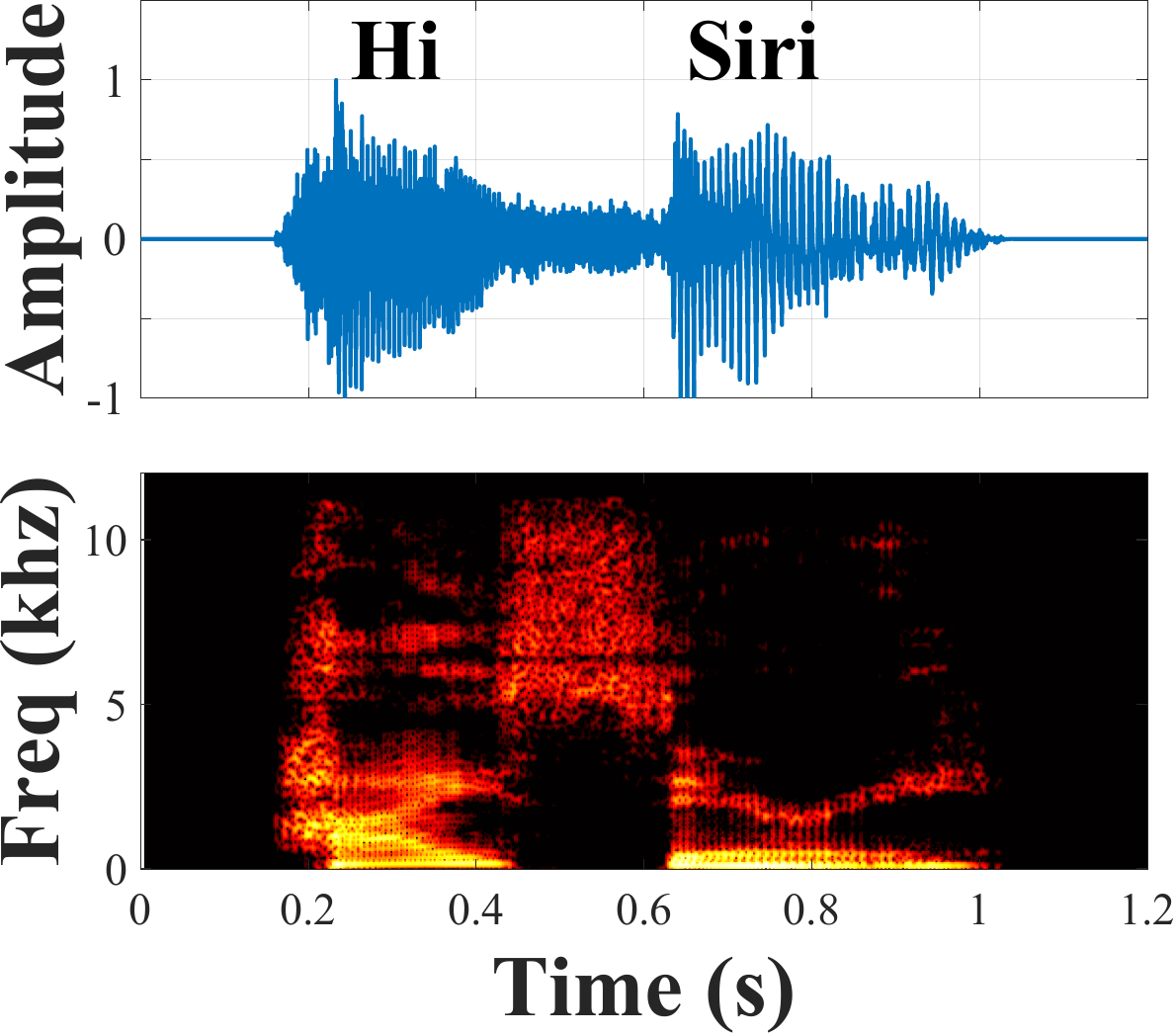}
    }
    \subfigure[Perturbation audio]{
        \centering
        \includegraphics[height = .27\linewidth]{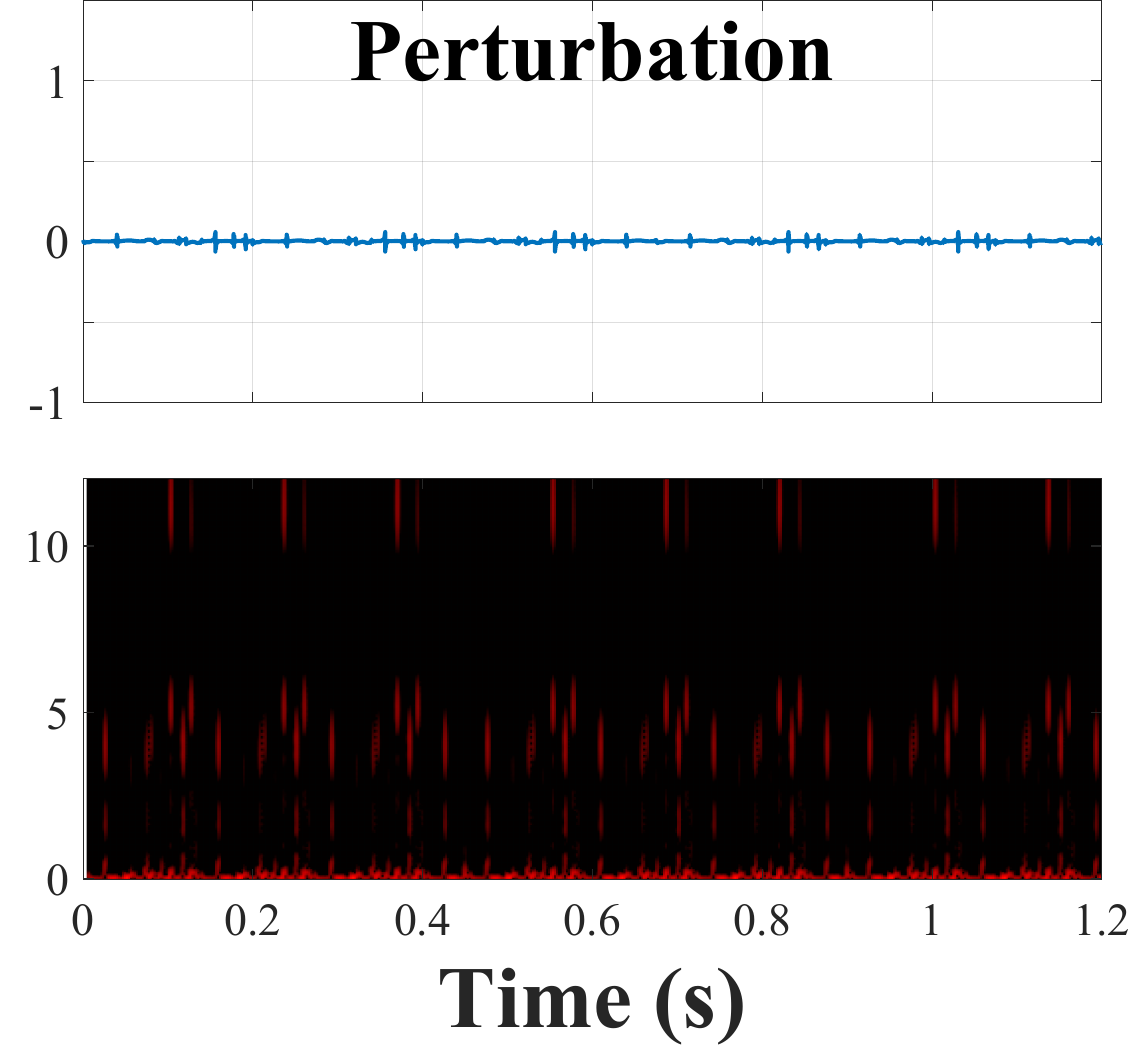}
    }
    \subfigure[Perturbed audio]{
        \centering
        \includegraphics[height = .27\linewidth]{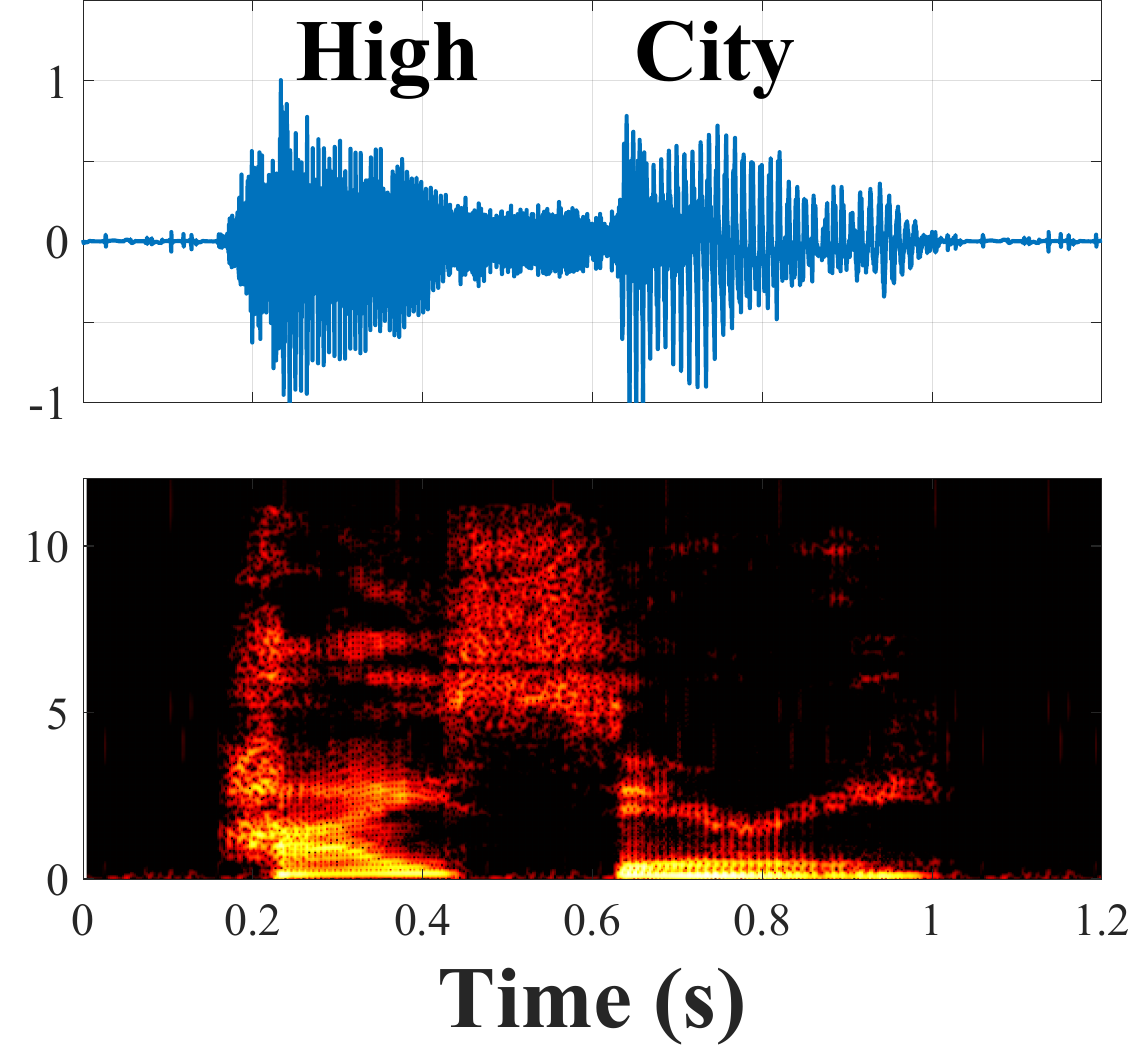}
    }
    \caption{\revc{Recognition results with/without perturbations}.}
    \label{fig:VAs_perturbations}
\end{figure}
\subsection{\revc{Why Only for Attack, Not for Defense?}}
\revc{Current IVAs primarily exploit the nonlinear effects of microphones, which arise from imperfections in microphone circuitry design. Specifically, when an input signal 
$S_{\text{in}}(t) = s_1(t) + s_2(t)$, 
where $s_1(t)$ is an amplitude-modulated voice carrier $v(t)\cos(2\pi f_c t)$ and 
$s_2(t)$ is a pure carrier $\cos(2\pi f_c t)$, 
is received by the microphone, the front-end circuitry performs nonlinear mixing on it, i.e.,}
\begin{align}
    S_{out}(t)=AS_{in}(t)+BS^2_{in}(t)+\cdots,
\end{align}
\revc{where $v(t)$ denotes the forged voice command signal, $f_c$ is the carrier frequency for ultrasonic modulation, and $A$ and $B$ are the gains of the linear and quadratic components, respectively.}

\revc{Additionally, due to the microphone's built-in low-pass filter, only the low-frequency components (i.e., $v(t)$) are retained, resulting in the following output signal:}
\begin{align}
    S_{out}(t)=\frac{B}{2}(v^2(t)+2v(t)+1)\approx v(t).
\end{align}
\revc{However, traditional defenses detect such attacks passively by identifying spectral features \cite{MicGuard2024}, i.e., frequency traces (e.g., straight lines) that persist across the entire temporal axis at a specific frequency. Although these methods can issue warnings, their countermeasures, such as disabling the voice assistant or emitting audible interfering signals, inevitably compromise VAs' normal functionality.}

Fortunately, methods such as the Fast Gradient Sign Method (FGSM) \cite{FGSM} and Projected Gradient Descent (PGD) \cite{PGD} have introduced Universal Adversarial Perturbation (UAP), which can generate universal perturbation signals to mislead recognition when added to normal voice, as shown in Fig.~\ref{fig:VAs_perturbations}.
However, perturbations generated by these methods not only interfere with normal speech but are also perceptible to humans. To address this, we aim to employ a universal perturbation signal $\delta(t)$, modulated onto a high-frequency carrier, to selectively block ultrasonic attack commands for active defense.

\revc{Remarkably, successful active defense must address two key challenges: (1) universal perturbation signals are unsynchronized with randomly transmitted attack signals, and (2) attack signals may be modulated at arbitrary carrier frequencies. The universal perturbation generation is detailed in Section~\ref{SUAD_defense}.}

\section{\revc{System Design}}
\revc{In this section, we first provide a brief overview of the \name. Then, we describe in detail the implementation of SUAD Attack and SUAD Defense.}

\begin{figure}[t]
    \centering
    \includegraphics[width=.8\linewidth]{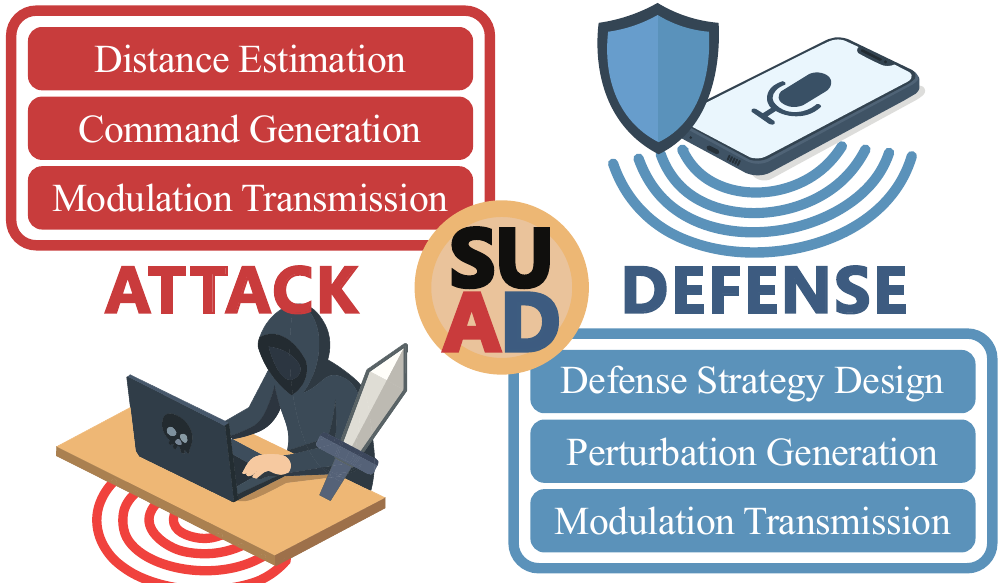}
    \caption{System architecture of SUAD Attack and Defense.}
    \label{fig:overview}
    \vspace{-1em}
\end{figure}
\subsection{System Overview of \name}
\revc{Considering real-world scenarios described in Section \ref{Threat_Model}, we propose SUAD, as illustrated in Fig.\,\ref{fig:overview}, which explores IVAs on VAs via solid channels, along with a universal defense against such attacks.}

\revc{For SUAD Attack, it allows Eve to attack VAs on Bob's smartphone by transmitting inaudible malicious commands through a solid medium. Specifically, to counteract the dispersion effect in solid-channel transmission, Eve first uses the MIC array to locate Bob's smartphone and estimate the attack distance by detecting the impact sound of its placement, i.e., a key contributor to signal dispersion. Subsequently, parameters such as the attack distance, Bob's voice \revr{(i.e., recording $\pm$\,5 seconds)}, and the recording transmitted through the table are simultaneously fed into a perceptual zero-sample speech cloning model with fused multi-head architecture, generating speech commands embedded with Bob's voiceprint features and inverse solid-channel interference. Ultimately, the generated commands are modulated onto a high-frequency carrier and transmitted through a solid medium to the victim's device via a piezoelectric transmitter.}

\revc{For SUAD Defense, an inaudible perturbation signal is continuously emitted through the speaker of Bob's smartphone, defending against IVAs without affecting the normal operation of VAs. We first generate a spectral- and delay-adaptive universal perturbation signal against attacks using a gradient descent algorithm. This signal is then modulated to a fixed ultrasonic frequency supported by the smartphone, producing an ultrasonic perturbation signal. As a result, it obstructs attack signals while preserving the effectiveness of Bob's voice captured by the microphone.}

\subsection{Long-range SUAD Attack Adapted to Solid-Channel States}\label{SUAD_Attack}
\revc{To successfully implement IVAs over solid channels, the primary challenge lies in addressing the dispersion effect on ultrasonic attack signals. Specifically, the dispersion phenomenon is manifested as different frequencies of acoustic waves propagating at different speeds in the solid medium, resulting in different frequency components having relative time delays, thus causing the signal waveform broadening and distortion. Therefore, the impact of attack distance and solid material, especially in long-range scenarios, cannot be ignored.}

\begin{figure}[b]
    \setlength\abovecaptionskip{0pt}
    \setlength{\subfigcapskip}{0pt}
    \subfigure[Solid-channel signals]{    
        \centering
        \includegraphics[height =.25\linewidth]{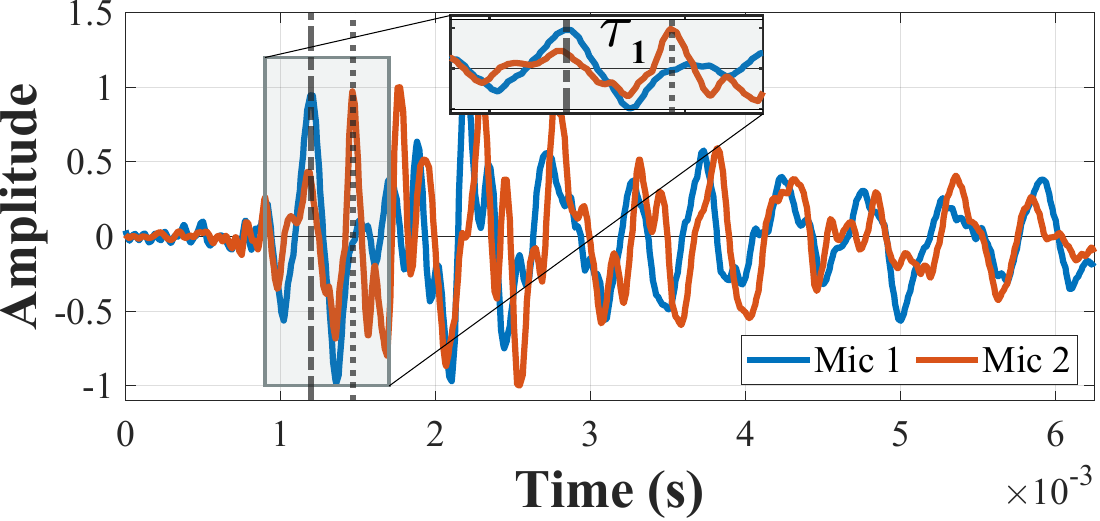}
        \label{location_a}
    }
    \subfigure[Localization via hyperbolas]{
        \centering
        \includegraphics[height = .23\linewidth]{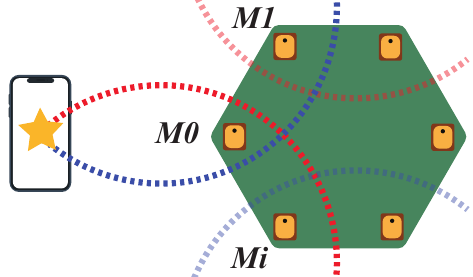}
        \label{location_b}
    }
    \caption{\revc{Attack distance estimation via 6 microphones' TDoA}.}
    \label{fig:location}
    \vspace{-1em}
\end{figure}

\subsubsection{\revc{Attack Distance Estimation}}
\label{Attack_Distance}
\revc{To weaken the effect of dispersion, SUAD Attack estimates the attack distance using TDOAs of microphones, which is then used to apply frequency-domain compensation to signals. We first detect the time $t_0$ when Bob places the smartphone by applying a threshold $\gamma$ to the short-duration energy envelope from microphone $\mathrm{M}_0$. Based on $t_0$, a 0.7\,ms segment is extracted as the solid-channel signal $S_0$, as shown in Fig.\,\ref{location_a}. Next, TDOA $\tau_i$ is obtained by locating the maximum of the cross-correlation between $S_0$ and the signal from microphone $\mathrm{M}_i$, where the distance difference is given by $\Delta d_i = \tau_i \cdot c_s$, with $i = 1, \dots, 5$, and $c_s$ denoting the speed of sound in solids. Assuming the impact point is $P = (x, y)$ and the position of $\mathrm{M}_i$ is $M_i=(x_i, y_i)$, with $M_0$ as the reference, the distance difference between $\mathrm{M}_i$ and $M_0$ is given by:}
\begin{equation}
    \|P - M_i\| - \|P - M_0\| = \Delta d_{i}.
\end{equation}
\revc{Then, as shown in Fig.\,\ref{location_b} the intersection of five hyperbolic curves derived from the six microphone positions yields the point $P$, from which the attack distance $L$ is calculated. Finally, we obtain propagation delays of different frequency components $\tau(f)=L/v(f)$, where $v(f)$ is the velocity in solids for frequency $f$.}
\begin{figure}[t]
    \centering
    \includegraphics[width =\linewidth]{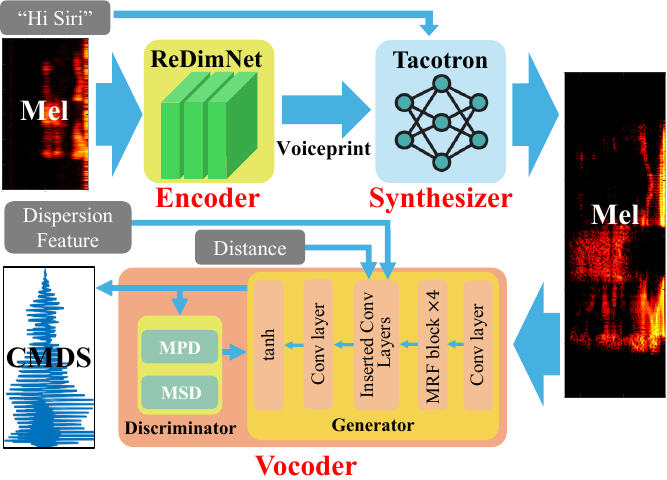}
    \caption{\revc{Model structure for attack command generation}.}
    \label{fig: model1}
    \vspace{-1em}
\end{figure}

\revc{Theoretically, the relative delay in the ultrasonic signal $s(t)$ propagating through a solid can be compensated, where $s(t)$ is given by:
\begin{equation}
    s(t) = \int_{-\infty}^{\infty} S(f)\, e^{j 2\pi f t} \, df,
\end{equation}
where $S(f)$ is the frequency domain signal. Subsequently,
we can obtain a compensated signal $s_{c}(t)$:
\begin{equation}
    s_{c}(t) = \int_{-\infty}^{\infty} S(f) \, e^{j 2\pi f (t - \tau(f))} \, df.
\end{equation}
By eliminating the frequency-dependent time delays caused by dispersion, the ultrasonic attack signal can release the original commands accurately at the target location.}

\subsubsection{Ultrasonic Attack Signal Generation}
However, even with an accurate distance, various complex factors (e.g., material properties) make it difficult to obtain $v(f)$, preventing direct compensation for dispersion.
\revc{
Furthermore, VAs typically require voiceprint authentication, meaning that attack commands must incorporate the user's voiceprint features. Therefore, SUAD Attack designs a multi-parameter modular attack command generation model, as shown in Fig.\,\ref{fig: model1}. It consists of three modules:  
i) \textit{Speech Encoder}, which extracts voiceprint features from Mel spectrograms of the user's recordings,  
ii) \textit{Synthesizer}, which converts textual commands into Mel spectrograms embedded with voiceprint features, and  
iii) \textit{Vocoder}, which fuses distance and material features into the Mel spectrograms to reconstruct the time-domain waveforms of the attack speech commands.}
 
The first module is implemented on ReDimNet \cite{redimnet}, which begins with a 2D convolutional layer to extract local time-frequency features from 2–3\,s Mel spectrograms of user recordings. \revr{Benefiting from this efficiency, only a short voice sample of approximately 5\,s is sufficient to construct a complete voiceprint.} Then, the extracted non-linguistic features, such as pitch and timbre, are fed into a Transformer to learn prosodic and articulatory patterns. Finally, the model's noise robustness is further enhanced by the self-attention mechanism \cite{vaswani2017attention}, which suppresses transient distortions caused by environmental disturbances and reverberation. The module can be pre-trained on the TIMIT speech dataset \cite{TIMIT}, and no parameter updates are required after training. It extracts the target user's vocal features as 256-dimensional embedding vectors for input into the synthesizer in the next stage. The synthesizer is a Tacotron-enhanced sequence-to-sequence model based on the Encoder-Attention-Decoder architecture \cite{tacotron2}, embedding the voiceprint vectors into the Mel spectrogram generated by the textual commands. The module is also a zero-shot model that requires no updates after a single training. Given command text and voiceprint features, it performs one-shot inference to generate the corresponding Mel spectrogram, ensuring high efficiency and real-time attack capability.

Finally, the vocoder serves as the core of SUAD Attack for eliminating dispersion effects. It adopts a HiFi-GAN \cite{HiFi-GAN} structure \revr{featuring a multi-receptive field fusion (MRF) mechanism} that can integrate features such as solid material and distance. This enables the conversion of the Mel spectrogram into a time-domain ultrasonic attack waveform embedded with inverse solid-channel interference to counteract dispersion. Specifically, we pre-collected an offline dataset of solid-channel propagation across various materials to train a dispersion feature extractor similar to an encoder. Using a multilayer structure, it captures dispersion features such as frequency distortion and delay patterns from the frequency domain. Then, we fuse HiFi-GAN with the dispersion feature vector and propagation distance (from Section\,\ref{Attack_Distance}) by \revr{inserting two convolutional layers after the MRF blocks and before the final Conv layer.} This enables the vocoder to adjust the output waveform based on material and distance. 

During training, the offline solid-channel dataset includes the same table materials as those used in scenarios.
After training, the pre-trained weights remain fixed. The attack waveform $A(t)$ can be generated by modifying only the material features in the fusion layer.
An ultrasonic carrier with frequency \(f_c=21\,\text{kHz}\) is then employed to modulate the waveform, producing an inaudible attack signal. This signal is transmitted through the table using a piezoelectric device to complete the attack on the VAs.

\subsection{Universal SUAD Defense without Affecting VAs }\label{SUAD_defense}
Successful IVAs ultimately rely on downconversion to inject attack commands into microphones. Such attacks can be effectively countered by jamming the injection commands. 
In particular, the defender can get various information from the speech recognition model. Therefore, unlike existing passive defenses \cite{Lipread_2018, Watchdog2020, MicGuard2024}, such as detecting anomalies in the speech spectrum, SUAD Defense introduces a UAP-based approach leveraging the smartphone’s speaker to emit inaudible perturbation signals for targeted protection against IVAs.

\begin{figure}[t]
    \subfigure[Spectrum after 2 random attacks]{    
        \centering
        \includegraphics[height =.245\linewidth]{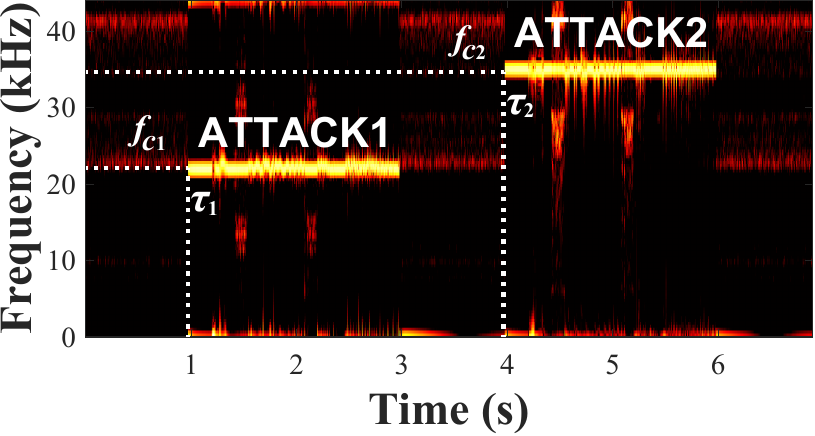}
    }
    \subfigure[Random attack time $\&$ frequency]{
        \centering
        \includegraphics[height = .245\linewidth]{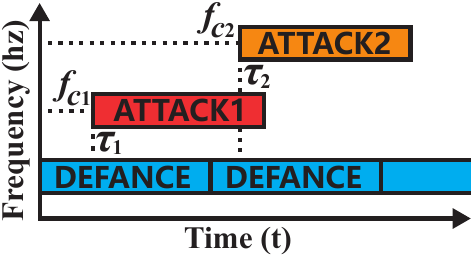}
    }
    \caption{\revc{Attacks with randomized time and carrier frequency}.}
    \label{fig:randomized}
\end{figure}

\subsubsection{\revc{Analysis of Ultrasonic Perturbations to IVAs}}
\revc{In practical scenarios, IVAs can be modulated to any inaudible frequency and send attack signals at any time, as shown in Fig. \ref{fig:randomized}. Consequently, the actual signal \( s_1'(t) \) received by the microphone can be modeled as a time-shifted version of the ideal signal \( s_1(t) \) due to a delay \( \tau \), and is expressed as:
\begin{align}
    s_1'(t) = s_1(t +\tau)=v(t+\tau)cos(2\pi f_c(t+\tau)),
\end{align}
where the content of the attack command  $v(t)$  and the carrier frequency $f_c$ may also change randomly, making IVAs difficult to defend against. Although the second carrier $s_2(t)$ of IVAs leaves a distinct and fixed feature (i.e., a straight line) in the spectrum, which can serve as an important basis for detection, this only helps potential victims to passively recognize such attacks. In most cases, victims become aware of the attack only after it has already been executed. Even if the attack is detected in time and defensive measures (e.g., turning off the microphone \cite{Watchdog2020}) are taken, these actions can interfere with the normal use of VAs.}

\revc{Therefore, we intend to generate a small-amplitude universal perturbation signal $\delta(t)$  in response to activation commands and modulate it into the inaudible band using a smartphone-supported frequency $f_x = 18\,\text{kHz} $. This ultrasonic perturbation signal is played cyclically. When an ultrasonic attack occurs, the signal captured by the microphone can be represented as:}

 \begin{small}
\begin{equation}
\begin{split}
&x(t+\tau)cos(2\pi f_c(t+\tau))+\delta(t)cos(2\pi f_x t)+cos(2\pi f_c(t+\tau))\\
    &\xrightarrow{\text{Downconversion}}{\text{}} x(t+\tau)+\delta^{'}(t),
\end{split}
\end{equation}
\end{small}
where $\delta^{'}(t)$ is the universal perturbation to the forged command $x(t+\tau)$. Specifically, the ultrasonic perturbation can leverage the second carrier of IVAs to down-convert it into a universal perturbation, thereby corrupting the integrity of the attack command to prevent it from activating VAs.

\revc{If only audible voice commands and ultrasonic perturbations are present, the received signal can be expressed as follows:}
\begin{align}
    v(t+\tau)+\delta(t)&cos(2\pi f_x t))\Rightarrow v(t).
\end{align}
\revc{The reason is that the ultrasonic perturbation functions merely as a small-amplitude, high-frequency background noise and does not interfere with the normal voice interaction of VAs.}

\subsubsection{\revc{Universal Perturbation Signal Generation}} 
\revc{To generate subtle, pervasive, and imperceptible universal perturbations \( \delta(t) \), SUAD Defense introduces a UAP-based \cite{UAP2019} method.}

\revc{First, we define a speech domain \( X = \left\{x_1(t), x_2(t), \ldots \right\} \), where \( x_i(t) \) represents speech signals that may be received by VAs on a user's smartphone, such as “Hi Siri”. Based on \( X \), we propose a defense model as follows:
}

\begin{align}
    \text{CER}(C(x_i(t)), C(x_i(t) + \delta'(t))) &> 0.7, & \forall x_i &\in X,
\end{align}

\revc{where \( C(\cdot) \) denotes a speech recognition model accessed by defender, and \( \text{CER}(x, y) \) represents the character error rate (CER), i.e., the edit distance \cite{Levenshtein_SPD66} between two recognition results. If perturbation signals are added to the speech, more than 70\% of the VA's results should be incorrect.}

\revc{Furthermore, since the attack signal may begin at any arbitrary point in time, the perturbation signal may not always be temporally aligned with the speech input \( x_i(t) \). To address this, we represent the speech input as \( x_i(t + \tau) \in [0, T] \), which indicates a circularly shifted result of \( x_i(t) \) by \( \tau \), aligned with the perturbation signal. For example, the original command “Hi Siri” may be transformed into “Siri Hi” during training to address uncertainty in the start time.}

\begin{algorithm}[t]
\setstretch{1.12}
\setlength\belowcaptionskip{-15pt}
\caption{ Time and Frequency Randomized UAP Training}
\label{algorithm1}
\KwIn{$X$: speech set, $T$: delay set, $F$: frequency set, $C$: speech recognition model, $f_x$: perturbation signal frequency, $\epsilon$: maximum magnitude, $N$: max iterations, $a$: regularization weight}
\KwOut{Universal perturbation $\delta(t)$}

Initialize $\delta(t) =\vec{0}$,\quad $r = 0$,\quad $k = 0$\;

\For{$x_i(t) \in X$}{

    Zero-pad $x_i(t)$ to fixed length\;
    
    \For{each $(\tau, f) \in T \times F$}{
    
        $\tilde{x}(t) \leftarrow x_i(t+\tau) \cdot \cos(2\pi f (t+\tau)) + \cos(2\pi f (t+\tau))$\;

        $\delta_{mod}(t) \leftarrow \delta(t) \cdot \cos(2\pi f_x t)$\;

        $x^{adv}(t) \leftarrow \text{Nonlinear}(\tilde{x}(t) + \delta_{mod}(t))$\;

        \While{$\text{CER}(C(x_i(t+\tau)), C(x^{adv}(t) + r)) \leq 0.7 \parallel k < N$}{
            Compute $\Delta\delta(t)$ minimizing:

            $a\|r\|^2 - \text{CTCLoss}(C(x^{adv}(t) + r), C(x_i(t+\tau)))$

            \textbf{subject to} $\|\delta(t) + r\|_\infty < \epsilon$\;

            Update: $\delta(t) \leftarrow \delta(t) + \Delta\delta(t)$, project to $\|\cdot\|_\infty < \epsilon$\;

            $k \leftarrow k + 1$\;
        }
    }
}
\Return $\delta(t)$;
\end{algorithm}

\revc{However, since the lengths of each \( x_i \) are not the same, we first normalize all \( x_i \) to the same length via zero-padding, before applying circular shifting. Finally, considering the impact of different carrier frequencies on the down-modulated universal perturbation, we modulate the above command signals using multiple carrier frequencies, forming the dataset used for model training. Thus, the constraints on the perturbation $\delta(t)$ can be summarized as follows:}
\begin{equation}
    \begin{split}
        \|\delta\|_\infty(t) &< \epsilon \\
    P_{x \sim X} (\text{CER}(x_i(t+\tau), C(x_i(t+\tau) + \delta'(t))) > 0.7) &\geq L,
    \end{split}
\end{equation}

\revc{where \( \epsilon \) denotes the maximum allowable magnitude in each iteration, which constrains the size of \( \delta(t)\), and \( L \) represents the defense success rate.
}

\revc{To solve the above constrained optimization problem, we design an iterative algorithm named Time and Frequency Randomized UAP Training, as presented in Algorithm~\ref{algorithm1}. Briefly, the entire iterative process proceeds by sequentially selecting \( x_i(t + \tau) \) and modulating it with different carrier frequencies \( f_x \). Then, the perturbation vector \( \delta(t) \cos(2\pi f_x t) \) is constructed step by step, and the down-conversion process is simulated to generate \( x_i(t + \tau) + \delta'(t) \). In each iteration, we compute the smallest perturbation increment \( \Delta\delta(t) \) and add it to the current perturbation \( \delta^{'}(t) \), resulting in a transcription error. Meanwhile, the perturbation amplitude is controlled by enforcing \( \|\delta\|_\infty < \epsilon \) to ensure it remains within an acceptable range. This process can be expressed as:}

\begin{equation}
\begin{split}
 &\Delta\delta_i^j \leftarrow \arg\min_r \|r\|_2 \quad \\
    &\text{s.t.} \quad \text{CER}(C(x_i(t+\tau)), C(x_i(t+\tau) + \delta'(t) + r)) > 0.7.
\end{split}
\end{equation}

\revc{This is equivalent to maximizing the loss between the predicted probability distribution of the perturbed result \( C(x_i(t + \tau) + \delta'(t) + r) \) and the original result \( C(x_i(t + \tau)) \), i.e.,}
\begin{equation}
\begin{split}
&\min_r a\|r\|^2 - \text{CTCLoss}(x_i(t+\tau) + \delta'(t) + r, C(x_i(t+\tau)))\\
&\text{s.t.} \quad \|\delta'(t) + r\|_\infty < \epsilon,
\end{split}
\end{equation}

\revc{where $\text{CTCLoss}()$ denotes the Connectionist Temporal Classification (CTC) loss function, a commonly used objective in modern end-to-end speech recognition models to quantify the difference between the model output and the original transcription after perturbations are applied. Since the problem is a non-convex optimization task, we approximate its solution using the Iterative Gradient Sign Method (IGSM), i.e.,}
\begin{equation}
\begin{split}
r_0 &= \vec{0}, \\
    r_{N+1} &= \text{cut}_{(x_i(t+\tau), \epsilon)} \{ r_N + a \cdot \text{sign} ( \nabla_{(x_i(t+\tau) + \delta'(t) + r)} \,\\ &\text{CTCLoss}(x_i(t+\tau) + \delta'(t) + r, C(x_i(t+\tau))) ) \},
\end{split}
\end{equation}
\revc{where \( \text{cut}() \) denotes a clipping function, analogous to $\text{clip}()$ in image-based attacks. It segments the perturbation according to the audio's sampling points to ensure that the updated perturbation amplitude does not exceed \( \epsilon \). The regularization parameter \( a \) denotes the update step size for each iteration, and its value is determined via hyperparameter search on the validation set to maximize the attack success rate under the constraint of the maximum allowable perturbation magnitude. 
The operator \( \text{sign}() \) indicates the sign of the gradient, specifying the direction (positive or negative) of the perturbation update. 
The term \( \nabla(x_i(t + \tau) + \delta'(t) + r) \) denotes the gradient computed with respect to the current perturbed input signal, identifying the direction in which the input is most sensitive to the $\text{CTCLoss}()$.}

\revc{Finally, we utilize the public dataset and the TTS model introduced in the previous section to generate the training and test sets. During training, the universal adversarial perturbation is updated using the gradient descent method until it successfully defends against any sample in the test set.}

\begin{figure}[t]
    \setlength\abovecaptionskip{-1pt}
    \setlength{\subfigcapskip}{1pt}
    \subfigure[Attack devices]{    
        \centering
        \includegraphics[height =.315\linewidth]{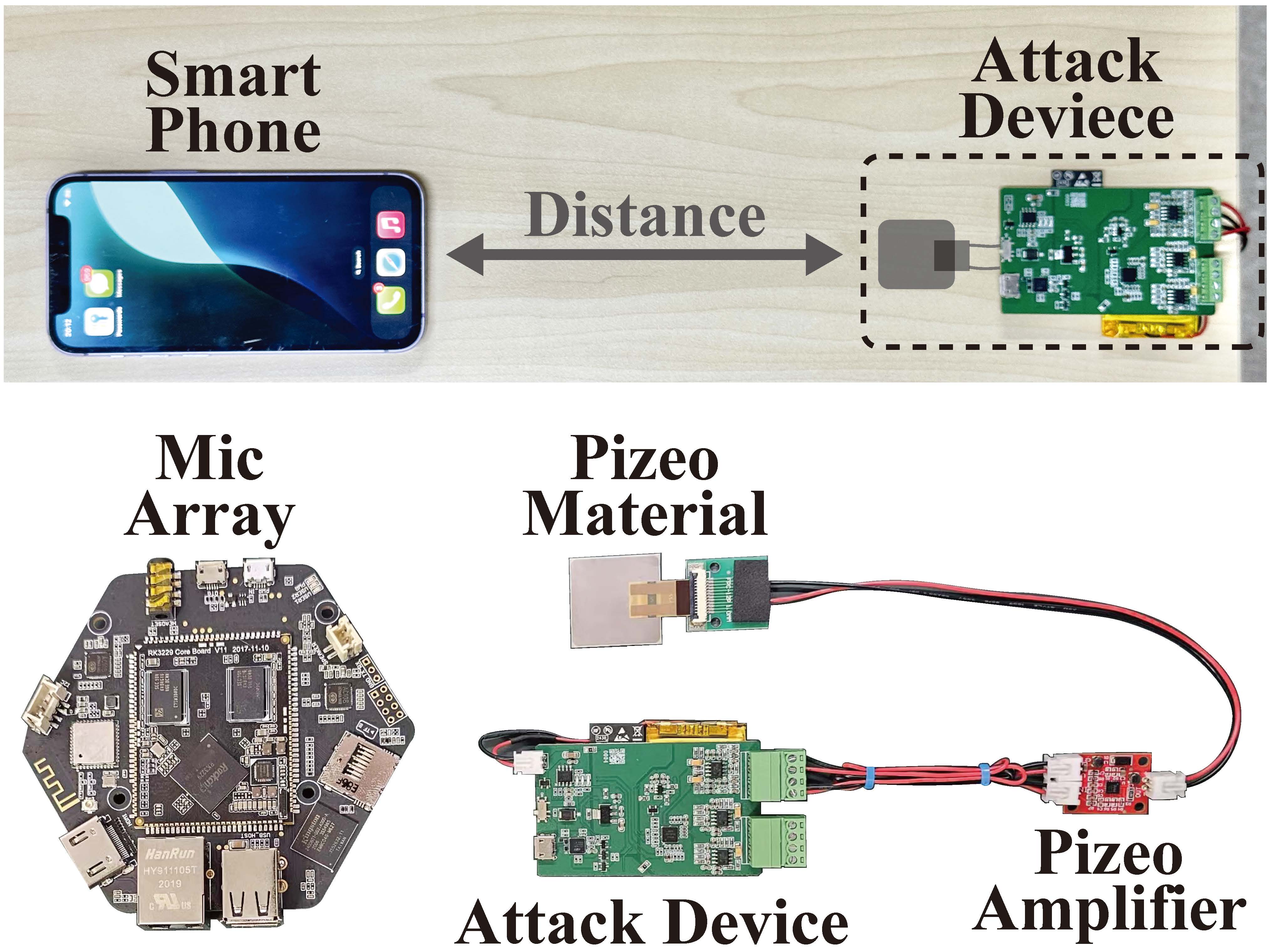}
        \label{fig:devices}
    }
    \hspace{-1em}
    \subfigure[Solid materials]{
        \centering
        \includegraphics[height = .315\linewidth]{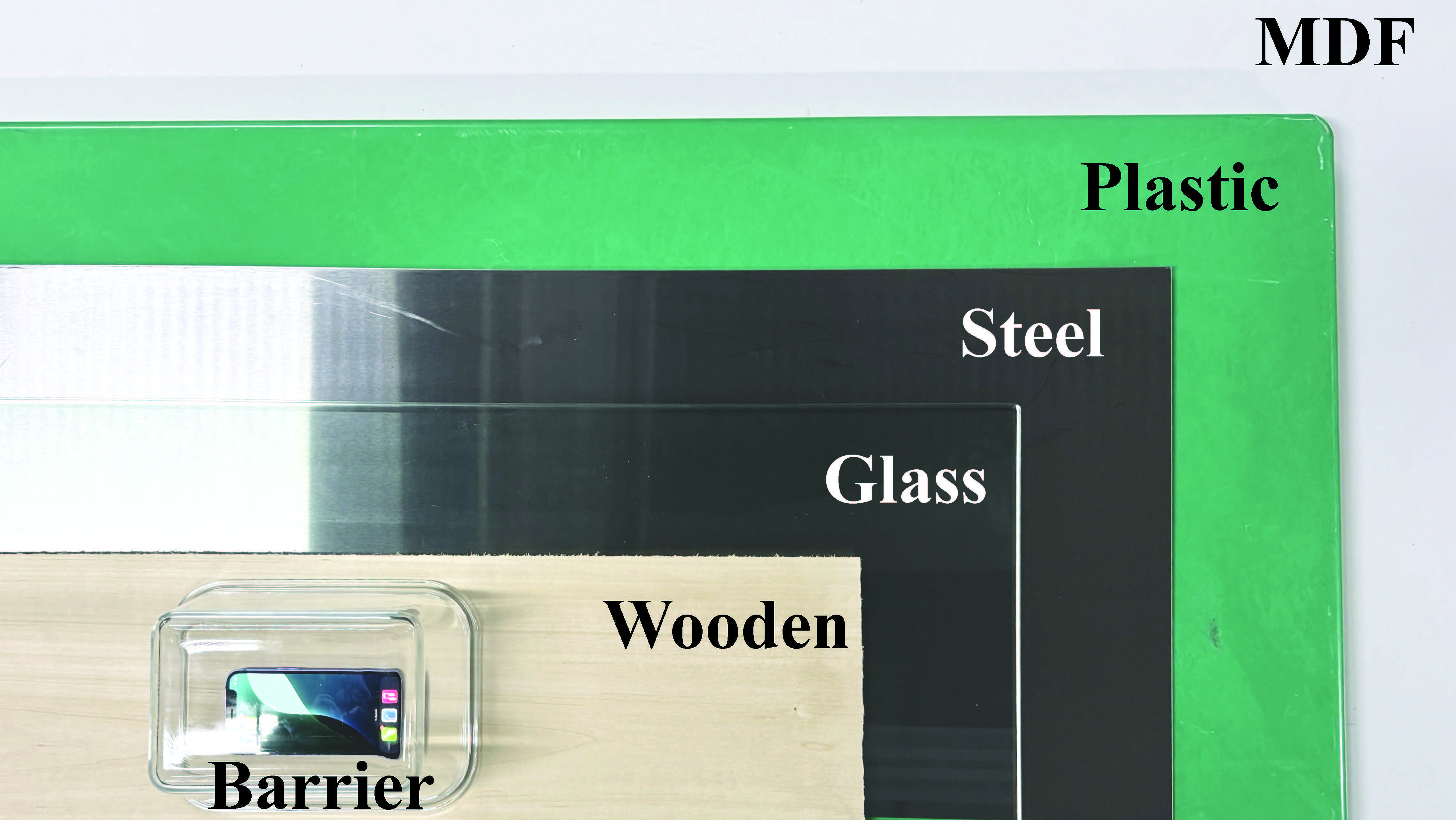}
        \label{fig:Materials}
    }
    \caption{\revc{Experimental Setup}.}
    \label{fig: setup}
    \vspace{-1em}
\end{figure}

\section{Evaluation}
\revc{In this section, we present the experimental setup and evaluate the overall performance of SUAD Attack and Defense, followed by an analysis of the impact of various factors.}
\subsection{Experimental Setup}
\subsubsection{Implementation}
\revc{As shown in Fig.~\ref{fig:devices}, we utilize a six-microphone array as the receiver and a piezoelectric transducer as the transmitter, both equipped with communication and power modules. The receiver sends audio recordings (i.e., the smartphone hitting) to a Linux server with an NVIDIA RTX 4090 GPU and an Intel Xeon Gold CPU for generating forged commands, which are sent to the transmitter for attacking VAs. For SUAD Defense, generated perturbation signals are modulated to 18~kHz for defending against IVAs.}
\subsubsection{Data Collection}
\revc{We recruited 5 subjects (3 males and 2 females), aged from 22 to 33 years old. Each subject was instructed to record activation and execution commands compatible with various VAs to construct a comprehensive voice command set, as well as an attack command set comprising 50 commonly used words. Each subject recorded approximately 5 minutes of voice commands. 3-7 word commands were designed to simulate realistic attack scenarios (e.g., sending SMS). Furthermore, we construct an audio dataset containing dispersion features by collecting audio propagating through solid media, such as materials shown in Fig.~\ref{fig:Materials}. }

\subsection{SUAD Attack Performance}

\begin{figure}[t]
    \centering
    \subfigure[VAs activation]{    
        \centering
        \includegraphics[width =.7\linewidth]{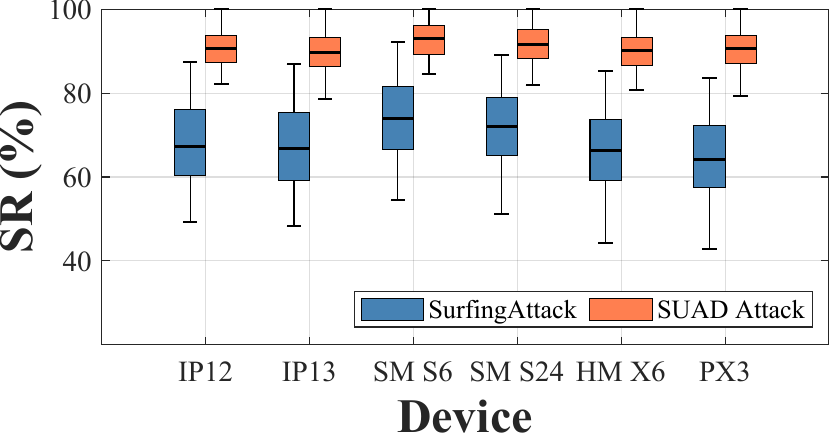}
        \label{fig:Attack_Overall1}
    }
    \subfigure[Command execution]{
        \centering
        \includegraphics[width = .7\linewidth]{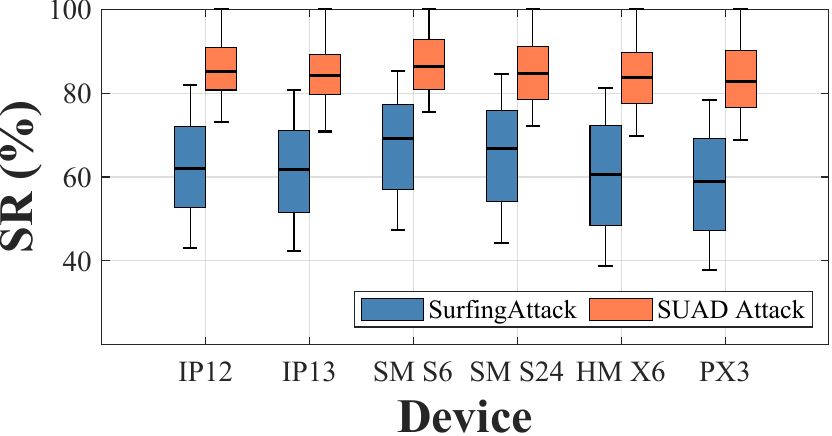}
        \label{fig:Attack_Overall2}
    }
    \caption{\revc{Overall performance of SUAD Attack}.}
    \label{fig:Attack_Overall}
\end{figure}

\begin{figure}[t]
    \centering
    \includegraphics[width=.7\linewidth]{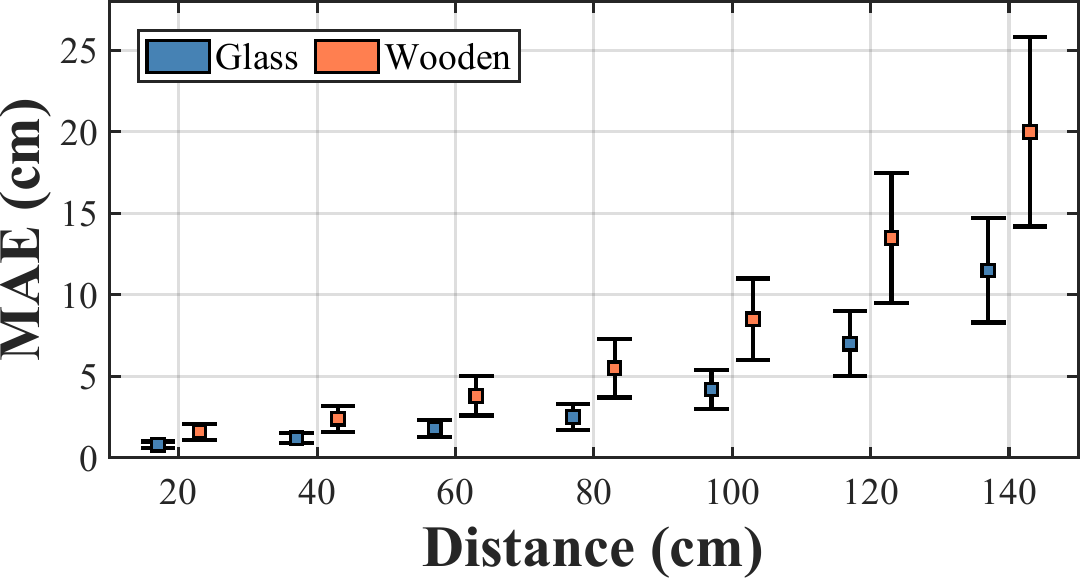}
    \caption{Distance estimation accuracy under different attack distances.}
    \label{fig:distance_estimate}
\end{figure}
\begin{figure}[t]
    \centering
    \includegraphics[width=.7\linewidth]{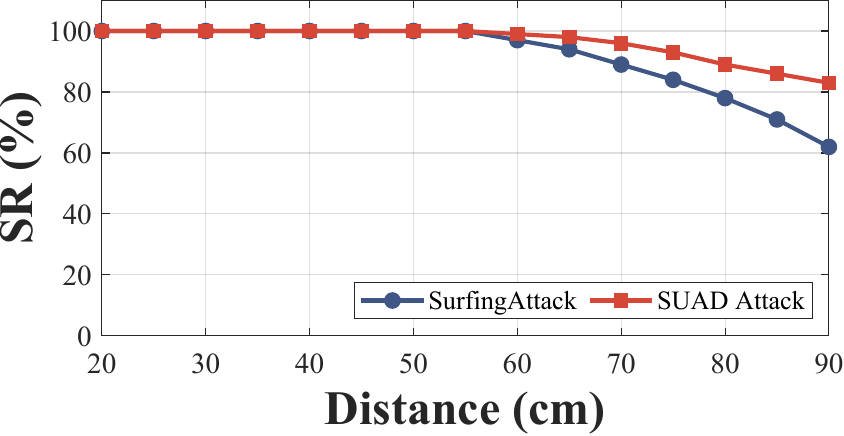}
    \caption{Impact of different distances on SUAD Attack performance.}
    \label{fig:Distance1}
\end{figure}

\begin{figure}[t]
    \centering
    \includegraphics[width=.7\linewidth]{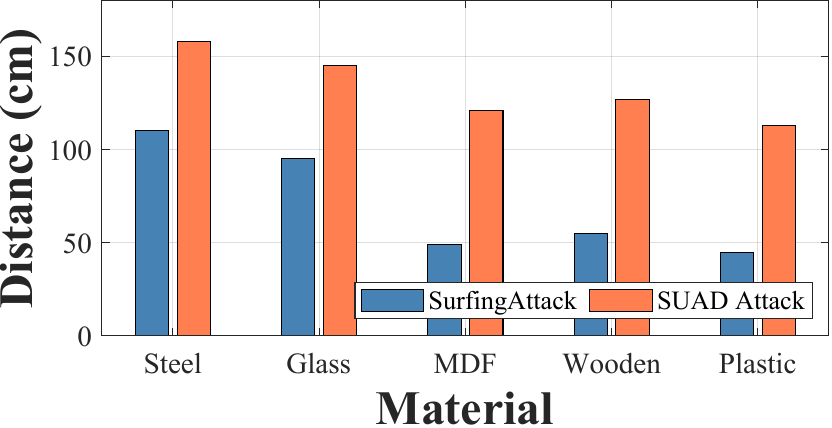}
    \caption{Impact of different materials on SUAD Attack performance.}
    \label{fig:materials}
\end{figure}

\begin{figure}[t]
    \centering
    \includegraphics[width=.7\linewidth]{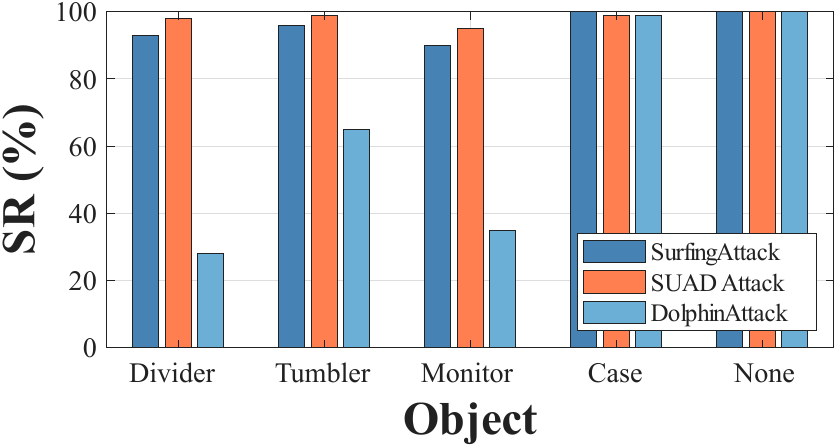}
    \caption{Impact of different objects on SUAD Attack performance.}
    \label{fig:Zhangai}
\end{figure}

\begin{figure}[t]
    \centering
    \includegraphics[width=.7\linewidth]{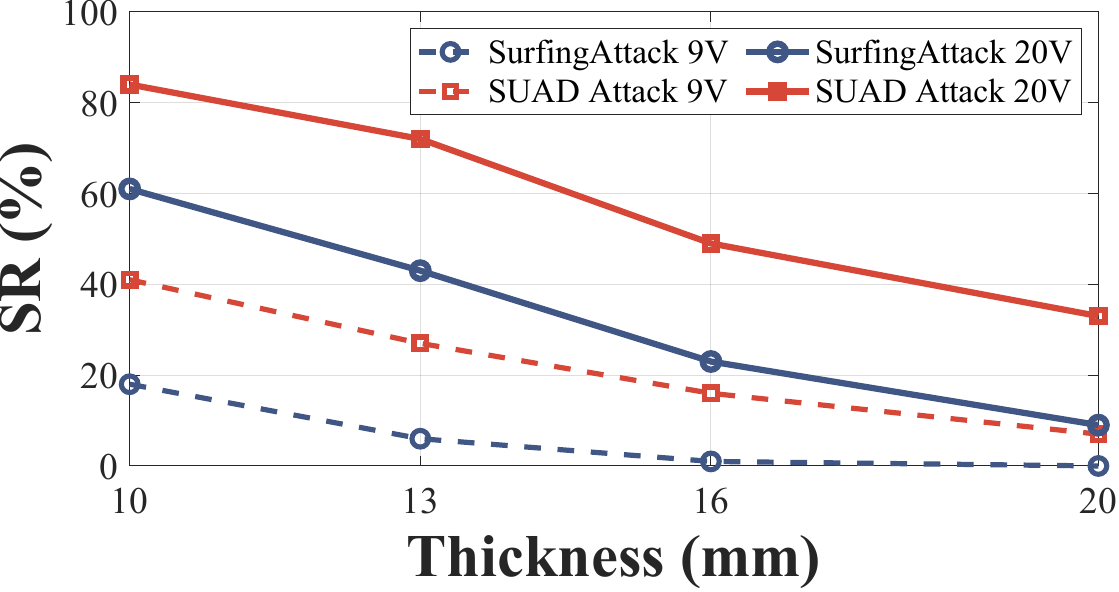}
    \caption{Impact of plate thickness under different voltages.}
    \label{fig:Thickness}
\end{figure}

\begin{figure}[t]
    \centering
    \includegraphics[width=.7\linewidth]{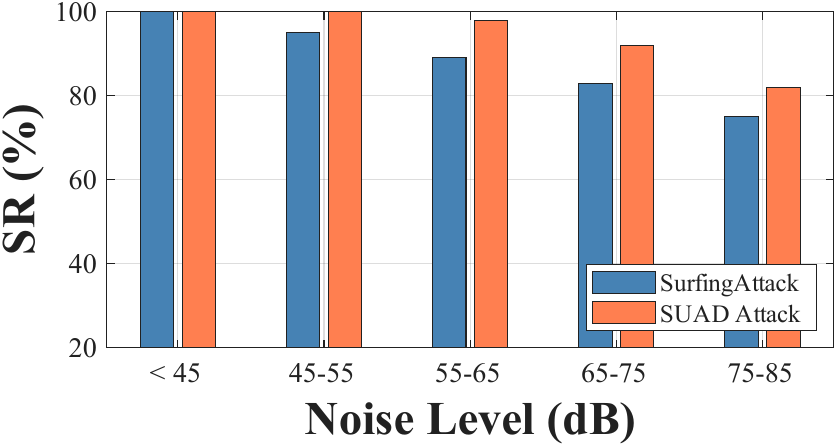}
    \caption{Impact of different noise levels on SUAD Attack performance.}
    \label{fig:Noise}
\end{figure}

\subsubsection{Localization Performance}
\revr{Before evaluating the overall performance of SUAD Attack, we assess the accuracy and stability of its localization module. The experiments adopt the same six-microphone array and receiver configuration as described in Section~\ref{SUAD_Attack}. The attack distance ranges from 20\,cm to 140\,cm, and each setting is repeated 10\,times on both glass and wooden tables to ensure consistency. We use the Mean Absolute Error (MAE) to evaluate localization accuracy, defined as the absolute difference between the estimated distance and the ground-truth value. As shown in Fig.\ref{fig:distance_estimate}, SUAD Attack maintains low localization errors across all tested distances, with the glass surface exhibiting superior performance. Specifically, for the glass material, the average MAE remains below 2\,cm when the distance is within 60\,cm, and gradually increases to approximately 11.5\,cm at 140\,cm. While the wooden table introduces slightly higher errors due to material properties, reaching 20.0\,cm at the maximum distance, these results demonstrate that the proposed TDoA-based localization method achieves stable and robust performance in solid-channel scenarios, providing reliable distance estimation for subsequent dispersion compensation.}

\subsubsection{Overall Attack Performance}
\revc{We evaluated the success rate (SR) of SUAD Attack in activating VAs using forged commands, comparing with SurfingAttack across multiple devices, including iPhone 12/13 (IP 12/13), Samsung S6/S24 (SM S6/S24), HUAWEI Mate X6 (HM x6), and Pixel 3 (PX3). As shown in Fig.~\ref{fig:Attack_Overall1}, SUAD Attack achieves a median activation success rate exceeding 89.8\% across different smartphones, demonstrating that its forged voice commands effectively simulate real user input. In comparison, SurfingAttack's median activation success rate does not exceed 74\%.
It may be attributed to dispersion effects in solid media that distort the transmitted signals, causing VAs to misidentify activation commands as other content. In addition, we evaluated commands such as 'turn on camera' and 'play music' to further evaluate the effectiveness of SUAD Attack. As shown in Fig.~\ref{fig:Attack_Overall2}, the median success rate of command execution for SUAD Attack is significantly lower than that for VAs activation in Fig.~\ref{fig:Attack_Overall1}, but remains above 82.8\%. Similarly, due to increased command length, which raises the misrecognition by VAs, SurfingAttack exhibits decreasing results. Notably, due to architectural, acoustic sensor, and VA consistency, both attacks exhibit similar performance on devices from the same manufacturer, such as iPhone and Samsung. Despite hardware variability, SUAD Attack consistently achieves high success rates, demonstrating its strong adaptability. }
\begin{figure}[t]
    \centering
    \includegraphics[width=.7\linewidth]{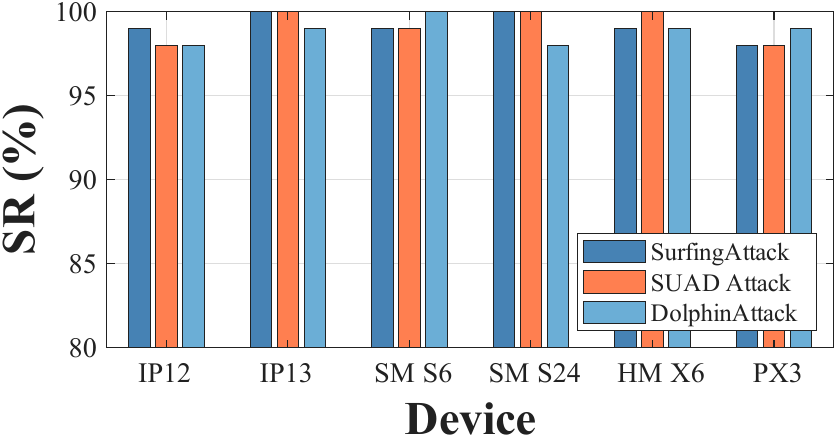}
    \caption{Defense performance against IVAs (DSR).}
    \label{fig:DSR}
\end{figure}

\begin{figure}[t]
    \centering
    \includegraphics[width=.7\linewidth]{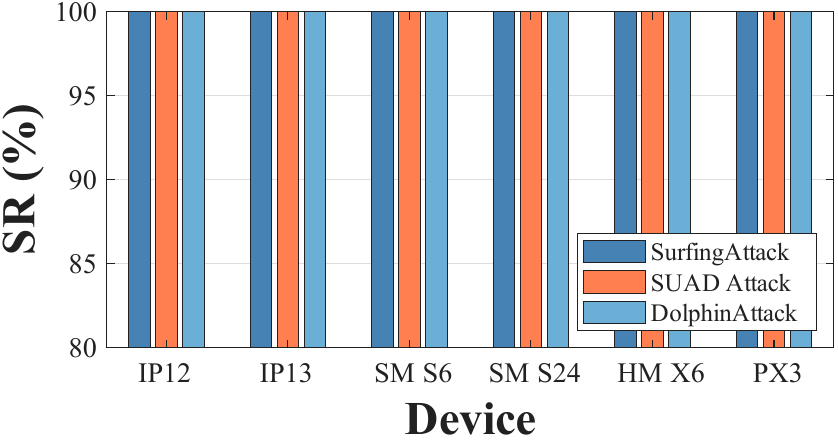}
    \caption{VAs activation performance under defense (AFR).}
    \label{fig:AFR}
\end{figure}

\begin{figure}[t]
    \centering
    \includegraphics[width=.7\linewidth]{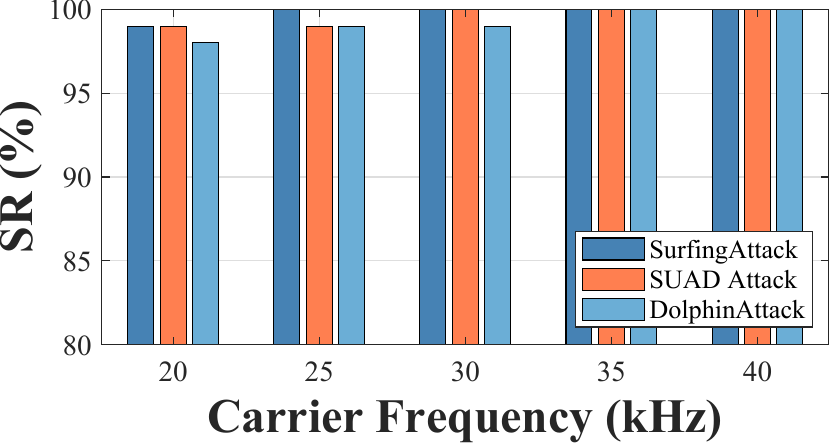}
    \caption{Impact of different frequencies on SUAD Defense performance.}
    \label{fig:Frequency}
\end{figure}

\subsubsection{Performance under Different Attack Distances}
\revc{Since victims may place their smartphones arbitrarily on the table, we further evaluated the impact of varying attack distances on SUAD Attack performance. Forged commands are transmitted at distances ranging from 20 to 90~cm, and the execution success rate is used as the evaluation metric. The results in Fig.~\ref{fig:Distance1} show that both SUAD Attack and SurfingAttack maintain a 100\% success rate at distances up to 55~cm. However, as the distance increases, the performance of SurfingAttack degrades significantly, whereas SUAD Attack demonstrates greater robustness to distance variations. This can be attributed to SUAD Attack’s integration of inverse solid-channel interference during command generation, which effectively mitigates signal distortion caused by distance variations.}

\subsubsection{Performance under Different Materials}
\revc{Equation~\eqref{eq1} demonstrates that variations in solid materials significantly affect signal dispersion during propagation. We selected five materials, including wooden, glass, steel, plastic, and MDF, as transmission media and transmitted commands to the target device. The maximum attack distance where the execution success rate reaches 50\% (i.e., effective attack distance) is used as the performance metric. As shown in Fig.~\ref{fig:materials}, SUAD Attack achieves effective attack distances of 158, 145, 121, 127, and 113~cm across the five materials, all substantially exceeding those of SurfingAttack, which reach only 110, 95, 49, 55, and 45~cm, respectively. In particular, plastic exhibits the greatest impact on both systems among the tested materials, primarily due to its higher signal attenuation. Additionally, SUAD Attack incorporates an anti-dispersion component in its command generation, resulting in smaller performance variations across different media than SurfingAttack.
}

\subsubsection{Performance under Different Plate Thickness}
\revr{Equation~\eqref{eq1} also demonstrates that plate thickness significantly affects signal attenuation and frequency-dependent dispersion in solid-channel propagation. Accordingly, we evaluated attack performance on wooden boards with thicknesses ranging from 10\,mm to 20\,mm under attack voltages of 9\,V and 20\,V. As shown in Fig.\,\ref{fig:Thickness}, under a 9\,V attack signal, the success rate decreases nonlinearly with increasing thickness. SUAD Attack retains a 41\% success rate at 10\,mm, whereas SurfingAttack degrades rapidly and becomes nearly ineffective beyond 16\,mm. Increasing the attack voltage to 20\,V improves robustness for both attacks. Under this configuration, SUAD Attack attains 84\% success at 10\,mm and retains approximately 33\% success at 20 mm, whereas SurfingAttack, despite benefiting from voltage boosting, achieves only 9\% success at 20\,mm. Compared with simple voltage boosting, the dispersion-compensation mechanism incorporated during the SUAD Attack training phase more effectively mitigates propagation-induced loss, enabling sustained performance even in thick-panel scenarios.}

\subsubsection{Performance under Different Objects on The Table} 
In practical scenarios, the LoS path between the attack device and the target device may be obstructed by objects on the table. On the one hand, these objects obstruct LoS propagation through the air channel. On the other hand, their physical contact with the table may impact the solid-channel transmission. To evaluate the impact of different barriers, we evaluated the execution success rates of SUAD Attack, SurfingAttack, and DolphinAttack at a fixed distance of 50~cm, respectively. The results, shown in Fig.~\ref{fig:Zhangai}, indicate that DolphinAttack, which relies on air-channel propagation, is highly sensitive to barriers. Its success rate exhibits a strong negative correlation with barrier size, i.e., smaller obstacles result in higher attack effectiveness. In contrast, other attacks relying on solid-channel propagation are minimally affected by barriers, demonstrating greater robustness. This suggests that such attacks can effectively bypass the blocking effects of objects on the table.
\subsubsection{Performance under Different Noise Levels} 
\revc{To evaluate SUAD Attack's performance under varying noise conditions, we simulated real-world scenarios by playing background noise at different intensity levels. We selected the iPhone 13 as the target device, set the attack distance to 50~cm, and transmitted 100 execution commands under each noise level. As shown in Fig.~\ref{fig:Noise}, the execution success rates of both SUAD Attack and SurfingAttack decline with increasing noise levels. Specifically, the success rate of SUAD Attack decreases from 100\% to 82\%, whereas SurfingAttack experiences a sharper decline from 100\% to 75\%. This is because SurfingAttack is influenced by both ambient noise and solid-channel dispersion, leading to greater distortion in the downconverted signal received by the microphone. In addition, the success rates of SUAD Attack in normal noise environments (e.g., lower 65~dB), which may be tolerable for victims, remain above 89\%, demonstrating strong robustness.
}

\subsection{SUAD Defense Performance}
\subsubsection{Overall defense Performance}
\revc{Our defense technique aims to disrupt the downconversion of attack commands by transmitting universal perturbation signals modulated into the ultrasonic band, thereby preventing the smartphone's VA from recognizing forged voice commands. To evaluate the performance of SUAD Defense, we continuously emit ultrasonic perturbation signals from the target device while executing SUAD Attack, SurfingAttack, and DolphinAttack, each transmitting ultrasonic commands intended to activate the voice assistant on the corresponding device. The defense success rate is calculated as the ratio of successfully blocked attacks to the total number of attack attempts. The results shown in Fig.~\ref{fig:DSR} demonstrate the performance of SUAD Defense against SUAD Attack, SurfingAttack, and DolphinAttack under normal noise conditions at an attack distance of 50 cm. The defense achieves a success rate of over 98\% for each type of inaudible voice attack, indicating that our system maintains a minimal false positive rate under operating conditions. }

\revc{In addition, we have evaluated the impact of SUAD Defense on normal voice commands when defending against ultrasonic attacks. In this experiment, SUAD Attack, SurfingAttack, or DolphinAttack is launched against the target device while ultrasonic perturbation signals are emitted as a defense mechanism. Activation voice commands are played 100 times, and the voice assistant’s activation success rate is measured to evaluate the impact of the defense on its functionality. The results are presented in Fig.~\ref{fig:AFR}, showing that SUAD Defense did not interfere with the control of VAs by normal voice commands, regardless of the attack method employed. The reason is that the perturbation signal of SUAD Defense is modulated within the inaudible frequency band, specifically disrupting the recognition of ultrasonic forged commands. In contrast, for normal audible speech, it manifests merely as low-amplitude high-frequency noise, exerting little to no impact on recognition accuracy. This result underscores a key advantage of SUAD Defense: it can effectively counter IVAs while preserving the normal functionality of VAs.}

\subsubsection{Defense against different-frequency attacks}
\revc{The fundamental principle of IVAs is to modulate attack commands into the inaudible frequency band, which are then down-converted into audible commands through the microphone's nonlinearity. However, any frequency within the inaudible band can serve as a carrier frequency. To evaluate the effectiveness of SUAD Defense against ultrasonic attacks at various frequencies, we modulate the attack commands from different IVAs using distinct carrier frequencies and conduct 100 attack attempts on the same phone for each frequency. The results are shown in Fig.~\ref{fig:Frequency}, where the defense success rate of SUAD Defense increases slightly as the carrier frequency increases. This may be attributed to the fact that lower-frequency attack signals undergo less dispersion in solid media, resulting in stronger signal integrity and revealing potential vulnerabilities of the model to such attacks. Nevertheless, SUAD Defense remains above 98\% for attack commands at different frequencies, demonstrating its strong robustness against variations in carrier frequency.}

\begin{figure}[t]
  \centering
  \includegraphics[width=0.75\linewidth]{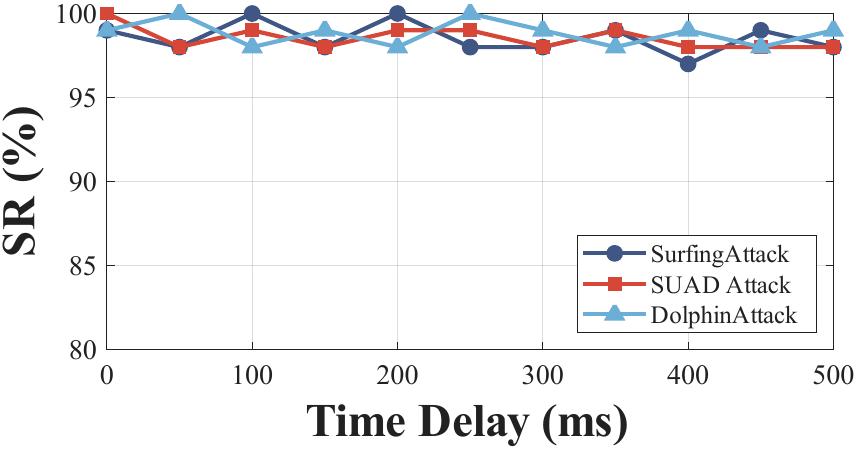}
  \caption{Impact of different time delays.}
  \label{fig:Delay}
\end{figure}

\subsubsection{Defense against randomly sent attacks}
\revc{Due to the random timing of attack signal transmissions, the perturbation signal may not be temporally synchronized with these signals. To evaluate the impact of time delay on SUAD Defense performance, we transmitted attack signals with varying delays relative to the perturbation signal using the same device. As shown in Fig.~\ref{fig:Delay}, while time delay introduces minor fluctuations in defense success rate, overall effectiveness remains largely unaffected. This robustness is attributed to the perturbation signal design, which accounts for time-delay-induced dispersion effects, and to adversarial learning employed during model training.}

\section{Related Work}
In this section, \revc{we first review various modalities of voice command injection attacks and their limitations, followed by an analysis of the shortcomings in existing defenses.}

\subsection{\revc{ Voice Command Injection Attacks}}
\revc{Although advances in AI have greatly contributed to the widespread application of VAs, it has also raised concerns about voice command injection attacks~\cite{Transduction_Shield, Inaudible_Attack2021, MagBackdoor, dai2023inducing, Ghost13, IEMI15, GhostTalk22, Light_Commands, zhang2017dolphinattack, roy2017backdoor, Lipread_2018, yan2020surfingattack, CapSpeaker_2021, spsm14_gvsattack, hidden2016, ndss19_adversarialattacks, usenix18_commandersong, survey21, Brother20}. For example, GVSAttack~\cite{spsm14_gvsattack} launches attacks on VAs by replaying voice commands. In addition, some studies attempt to conceal voice commands. Hidden Voice Commands~\cite{hidden2016} and Adversarial Attacks~\cite{ndss19_adversarialattacks} transform voice commands into white noise, and CommanderSong~\cite{usenix18_commandersong} embeds them into songs. However, such audible voice attacks are easily identified and interrupted by users. }

\revc{To hide attacks, researchers have explored various techniques for inaudible voice command injection~\cite{Transduction_Shield, Inaudible_Attack2021, MagBackdoor, dai2023inducing, Ghost13, IEMI15, GhostTalk22, Light_Commands, zhang2017dolphinattack, roy2017backdoor, Lipread_2018, yan2020surfingattack, CapSpeaker_2021}. For instance, MagBackdoor~\cite{MagBackdoor} and GhostTalk~\cite{GhostTalk22} leverage external magnetic fields to induce the target device's speaker to emit voice commands, but their effective range is limited to less than 6\,cm. Light Commands~\cite{Light_Commands} encodes speech information into laser beams, enabling attacks over distances up to 110\,m. However, it requires expensive devices and LoS. 
Consequently, ultrasound-based attacks (i.e., IVAs)~\cite{zhang2017dolphinattack, roy2017backdoor, Lipread_2018, yan2020surfingattack, CapSpeaker_2021} have attracted increasing attention due to their potential in IVAs. DolphinAttack~\cite{zhang2017dolphinattack} and Backdoor~\cite{roy2017backdoor} utilize nonlinear effects to modulate voice commands into high-frequency carriers for attacks. However, these approaches typically require an unobstructed LoS. Recently, SurfingAttack~\cite{yan2020surfingattack} has investigated the feasibility of performing ultrasonic injection through solid media, but ignores the interference of solid channel signals. Notably, SUAD Attack not only utilizes solid channels to penetrate physical barriers but also resists the dispersion effect of both distance and solid materials, thereby enabling long-range, interference-free attacks.}

\subsection{Defense against IVAs}
\revc{In recent years, various methods \cite{DualGuard20, Cacher19, zhou2019hidden, VAuth17, guan2023trustworthy,EarArray,RobustDetection21,arrayID22,zhang2017dolphinattack, Lipread_2018, MicGuard2024, Watchdog2020 } have been proposed to defend against IVAs. Some methods~\cite{DualGuard20, Cacher19} detect IVAs based on anomalies in the spectrum or sound field. For example, Cacher~\cite{Cacher19} distinguishes forged voice commands by establishing a biometric basis, Fieldprint, which represents sound field features. However, their effectiveness against ultrasonic band attacks remains unsatisfactory. Additionally, some multimodal methods~\cite{zhou2019hidden, VAuth17, guan2023trustworthy} attempt to utilize physiological or behavioral features to determine the authenticity of voice commands. Among them, MFF~\cite{guan2023trustworthy} detects attack attempts by leveraging synchronized video and audio data, but it requires high-precision sensors to ensure recognition accuracy. EarArray \cite{EarArray}, RobustDetection \cite{RobustDetection21}, and arrayID \cite{arrayID22} utilize microphone arrays for delay analysis and orientation analysis to extract spatial features during speech, but these techniques require dedicated hardware support. In contrast, modulation-based detection methods \cite{zhang2017dolphinattack, Lipread_2018, MicGuard2024, Watchdog2020} offer greater flexibility and can be applied to smartphones. For example, LipRead \cite{Lipread_2018} identifies IVAs by exploiting differences in frequency patterns between normal speech and fixed-frequency modulation attacks. However, these defenses function purely as passive detection mechanisms and do not actively prevent IVAs. In particular, these methods may interfere with the user's normal use of VAs, e.g., the system may disable the microphone after detecting an attack.
}

\section{Conclusion}
This paper presented SUAD that explores the feasibility of solid-channel attacks and universal defenses with inaudible ultrasonic signals. SUAD Attack presented a novel notion that solid-channel propagation can introduce significant distortions to acoustic signals. Thus, SUAD Attack employed a multi-parameter modular command generation model that adapts to solid channels by parameterizing attack distance, victim audio, and dispersion features. Additionally, SUAD Defense developed a universal defense against IVAs based on microphone nonlinearity, but without affecting VAs. An innovative UAP training method was designed to generate perturbation signals capable of blocking IVAs randomly sent at arbitrary frequencies. Extensive experiments demonstrate the effectiveness SUAD Attack and Defense.

\bibliographystyle{IEEEtran}
\bibliography{reference}

@inproceedings{roy2017backdoor,
  title={Backdoor: Making microphones hear inaudible sounds},
  author={Roy, Nirupam and Hassanieh, Haitham and Roy Choudhury, Romit},
  booktitle={Proceedings of the 15th Annual International Conference on Mobile Systems, Applications, and Services},
  pages={2--14},
  year={2017}
}

@inproceedings{zhang2017dolphinattack,
  title={Dolphinattack: Inaudible voice commands},
  author={Zhang, Guoming and Yan, Chen and Ji, Xiaoyu and Zhang, Tianchen and Zhang, Taimin and Xu, Wenyuan},
  booktitle={Proc. of the 24th ACM CCS},
  pages={103--117},
  year={2017}
}

@article{zhou2019hidden,
  title={Hidden voice commands: Attacks and defenses on the VCS of autonomous driving cars},
  author={Zhou, Man and Qin, Zhan and Lin, Xiu and Hu, Shengshan and Wang, Qian and Ren, Kui},
  journal={IEEE Wireless Communications},
  volume={26},
  number={5},
  pages={128--133},
  year={2019},
}

@inproceedings{yan2020surfingattack,
  author       = {Qiben Yan and
                  Kehai Liu and
                  Qin Zhou and
                  Hanqing Guo and
                  Ning Zhang},
  title        = {SurfingAttack: Interactive Hidden Attack on Voice Assistants Using
                  Ultrasonic Guided Waves},
  booktitle    = {27th Annual Network and Distributed System Security Symposium},
  year         = {2020}
}

@article{guan2023trustworthy,
  title={Trustworthy sensor fusion against inaudible command attacks in advanced driver-assistance systems},
  author={Guan, Jiwei and Pan, Lei and Wang, Chen and Yu, Shui and Gao, Longxiang and Zheng, Xi},
  journal={IEEE Internet of Things Journal},
  volume={10},
  number={19},
  pages={17254--17264},
  year={2023}
}

@inproceedings{EarArray,
  author       = {Guoming Zhang and
                  Xiaoyu Ji and
                  Xinfeng Li and
                  Gang Qu and
                  Wenyuan Xu},
  title        = {EarArray: Defending against DolphinAttack via Acoustic Attenuation},
  booktitle    = {28th Annual Network and Distributed System Security Symposium},
  year         = {2021}
}

@inproceedings{dai2023inducing,
  title={Inducing wireless chargers to voice out for inaudible command attacks},
  author={Dai, Donghui and An, Zhenlin and Yang, Lei},
  booktitle={IEEE symposium on security and privacy (SP)},
  pages={1789--1806},
  year={2023}
}

@inproceedings{hidden2016,
  author       = {Nicholas Carlini and
                  Pratyush Mishra and
                  Tavish Vaidya and
                  Yuankai Zhang and
                  Micah Sherr and
                  Clay Shields and
                  David A. Wagner and
                  Wenchao Zhou},
  title        = {Hidden Voice Commands},
  booktitle    = {25th {USENIX} Security Symposium, {USENIX} Security},
  pages        = {513--530},
  year         = {2016}
}

@inproceedings{Light_Commands,
  author       = {Takeshi Sugawara and
                  Benjamin Cyr and
                  Sara Rampazzi and
                  Daniel Genkin and
                  Kevin Fu},
  title        = {Light Commands: Laser-Based Audio Injection Attacks on Voice-Controllable
                  Systems},
  booktitle    = {29th {USENIX} Security Symposium},
  pages        = {2631--2648},
  year         = {2020}
}

@inproceedings{Lipread_2018,
  author       = {Nirupam Roy and
                  Sheng Shen and
                  Haitham Hassanieh and
                  Romit Roy Choudhury},
  title        = {Inaudible Voice Commands: The Long-Range Attack and Defense},
  booktitle    = {15th {USENIX} Symposium on Networked Systems Design and Implementation},
  pages        = {547--560},
  year         = {2018}
}

@article{vaswani2017attention,
  title={Attention is all you need},
  author={Vaswani, Ashish and Shazeer, Noam and Parmar, Niki and Uszkoreit, Jakob and Jones, Llion and Gomez, Aidan N and Kaiser, {\L}ukasz and Polosukhin, Illia},
  journal={Advances in neural information processing systems},
  volume={30},
  year={2017}
}

@article{Voice_Payment,
  author       = {Jingjin Li and
                  Chao Chen and
                  Mostafa Rahimi Azghadi and
                  Hossein Ghodosi and
                  Lei Pan and
                  Jun Zhang},
  title        = {Security and privacy problems in voice assistant applications: {A} survey},
  journal      = {Comput. Secur.},
  volume       = {134},
  pages        = {103448},
  year         = {2023}
}

@INPROCEEDINGS{VAs,
  author={Garai, Pooja and Nikam, Aditya and Kerkar, Suyash and Deshmukh, Samruddhi and Mane, Atharva and Bhise, Suvarna},
  booktitle={IITCEE}, 
  title={Voice AI-Intelligence Based Voice Assistant}, 
  year={2025},
  volume={},
  number={},
  pages={1-6}
  }

@INPROCEEDINGS{Siri,
  author={Jampala, Rahul and Kola, Devisri Santosh and Gummadi, Adithya Nagendra and Bhavanam, Meghana and Rani Pannerselvam, Ithaya},
  booktitle={IDCIoT}, 
  title={The Evolution of Voice Assistants: From Text-to-Speech to Conversational AI}, 
  year={2024},
  pages={1332-1338}
}

@INPROCEEDINGS{MagBackdoor,
  author={Liu, Tiantian and Lin, Feng and Wang, Zhangsen and Wang, Chao and Ba, Zhongjie and Lu, Li and Xu, Wenyao and Ren, Kui},
  booktitle={2023 IEEE Symposium on Security and Privacy (SP)}, 
  title={MagBackdoor: Beware of Your Loudspeaker as A Backdoor For Magnetic Injection Attacks}, 
  year={2023},
  volume={},
  number={},
  pages={3416-3431}
}

@inproceedings{Transduction_Shield,
  author       = {Yazhou Tu and
                  Vijay Srinivas Tida and
                  Zhongqi Pan and
                  Xiali Hei},
  title        = {Transduction Shield: {A} Low-Complexity Method to Detect and Correct
                  the Effects of {EMI} Injection Attacks on Sensors},
  booktitle    = {Proc. of the 2021 ACM Asia CCS},
  pages        = {901--915},
  year         = {2021}
}

@ARTICLE{Inaudible_Attack2021,
  author={Xu, Zhifei and Hua, Runbing and Juang, Jack and Xia, Shengxuan and Fan, Jun and Hwang, Chulsoon},
  journal={IEEE Transactions on Microwave Theory and Techniques}, 
  title={Inaudible Attack on Smart Speakers With Intentional Electromagnetic Interference}, 
  year={2021},
  volume={69},
  number={5},
  pages={2642-2650}
}

@inproceedings{CapSpeaker_2021,
  author       = {Xiaoyu Ji and
                  Juchuan Zhang and
                  Shui Jiang and
                  Jishen Li and
                  Wenyuan Xu},
  title        = {CapSpeaker: Injecting Voices to Microphones via Capacitors},
  booktitle    = {In Proc. of the 28th {ACM CCS}},
  pages        = {1915--1929},
  year         = {2021}
}

@article{dispersion,
author = {Vladimir I. Erofeev and Alexey Malkhanov},
title = {Nonlinear Acoustic Waves in Solids with Dislocations},
journal = {Procedia IUTAM},
volume = {23},
pages = {228-235},
year = {2017}
}

@inproceedings{redimnet,
  author       = {Ivan Yakovlev and
                  Rostislav Makarov and
                  Andrei Balykin and
                  Pavel Malov and
                  Anton Okhotnikov and
                  Nikita Torgashov},
  title        = {Reshape Dimensions Network for Speaker Recognition},
  booktitle    = {25th Annual Conference of the International Speech Communication Association},
  year         = {2024}
}

@article{TIMIT,
author = {Garofolo, J. and Lamel, Lori and Fisher, W. and Fiscus, Jonathan and Pallett, D. and Dahlgren, N. and Zue, V.},
pages = {},
title = {TIMIT Acoustic-phonetic Continuous Speech Corpus},
journal = {Linguistic Data Consortium},
year = {1992}
}

@inproceedings{tacotron2,
  author       = {Jonathan Shen and
                  Ruoming Pang and
                  Ron J. Weiss and
                  Mike Schuster and
                  Navdeep Jaitly and
                  Zongheng Yang and
                  Zhifeng Chen and
                  Yu Zhang and
                  Yuxuan Wang and
                  R. J. Skerry{-}Ryan and
                  Rif A. Saurous and
                  Yannis Agiomyrgiannakis and
                  Yonghui Wu},
  title        = {Natural {TTS} Synthesis by Conditioning Wavenet on {MEL} Spectrogram
                  Predictions},
  booktitle    = {IEEE ICASSP},
  pages        = {4779--4783},
  year         = {2018}
}

@inproceedings{HiFi-GAN,
  author       = {Jungil Kong and
                  Jaehyeon Kim and
                  Jaekyoung Bae},
  title        = {HiFi-GAN: Generative Adversarial Networks for Efficient and High Fidelity
                  Speech Synthesis},
  booktitle    = { NeurIPS },
  year         = {2020}
}

@article{Watchdog2020,
  author       = {Jian Mao and
                  Shishi Zhu and
                  Xuan Dai and
                  Qixiao Lin and
                  Jianwei Liu},
  title        = {Watchdog: Detecting Ultrasonic-Based Inaudible Voice Attacks to Smart Home Systems},
  journal      = {{IEEE} Internet Things J.},
  volume       = {7},
  number       = {9},
  pages        = {8025--8035},
  year         = {2020}
}

@inproceedings{MicGuard2024,
  author       = {Tiantian Liu and
                  Feng Lin and
                  Zhongjie Ba and
                  Li Lu and
                  Zhan Qin and
                  Kui Ren},
  title        = {MicGuard: {A} Comprehensive Detection System against Out-of-band Injection
                  Attacks for Different Level Microphone-based Devices},
  booktitle    = {Proc. of the 33rd USENIX Conference
on Security Symposium},
pages={3963–3978},
  year         = {2024}
}

@inproceedings{UAP2019,
  author       = {Paarth Neekhara and
                  Shehzeen Hussain and
                  Prakhar Pandey and
                  Shlomo Dubnov and
                  Julian J. McAuley and
                  Farinaz Koushanfar},
  title        = {Universal Adversarial Perturbations for Speech Recognition Systems},
  booktitle    = {20th Annual Conference of the International Speech Communication Association},
  pages        = {481--485},
  year         = {2019}
}

@article{Levenshtein_SPD66,
  author = {Levenshtein, Vladimir Iosifovich},
  title = {Binary codes capable of correcting deletions, insertions and reversals.},
  journal = {Soviet Physics Doklady},
  number = 8,
  pages = {707--710},
  volume = 10,
  year = 1966
}

@inproceedings{FGSM,
  author       = {Ian J. Goodfellow and
                  Jonathon Shlens and
                  Christian Szegedy},
  title        = {Explaining and Harnessing Adversarial Examples},
  booktitle    = {3rd International Conference on Learning Representations},
  year         = {2015}
}

@inproceedings{PGD,
  author       = {Aleksander Madry and
                  Aleksandar Makelov and
                  Ludwig Schmidt and
                  Dimitris Tsipras and
                  Adrian Vladu},
  title        = {Towards Deep Learning Models Resistant to Adversarial Attacks},
  booktitle    = {6th International Conference on Learning Representations},
  year         = {2018}
}

@inproceedings{AI_survey,
  author       = {Jaime Sevilla and
                  Lennart Heim and
                  Anson Ho and
                  Tamay Besiroglu and
                  Marius Hobbhahn and
                  Pablo Villalobos},
  title        = {Compute Trends Across Three Eras of Machine Learning},
  booktitle    = {IEEE IJCNN},
  pages        = {1--8},
  year         = {2022}
}

@inproceedings{spsm14_gvsattack,
    author      =   {Diao, Wenrui and Liu, Xiangyu and Zhou, Zhe and Zhang, Kehuan},
    title       =   {Your Voice Assistant is Mine: How to Abuse Speakers to Steal Information and Control Your Phone},
    booktitle   =   {Proc. of the 4th ACM Workshop on Security and Privacy in Smartphones \& Mobile Devices},
    pages       =   {63--74},
    year        =   {2014},
}

@inproceedings{usenix18_commandersong,
    author      =   {Yuan, Xuejing and Chen, Yuxuan and Zhao, Yue and Long, Yunhui and Liu, Xiaokang and Chen, Kai and Zhang, Shengzhi and Huang, Heqing and Wang, XiaoFeng and Gunter, Carl A.},
    title       =   {Commandersong: a systematic approach for practical adversarial voice recognition},
    year        =   {2018},
    booktitle   =   {Proc. of the 27th USENIX Conference on Security Symposium},
    pages       =   {49–-64},
}

@inproceedings{ndss19_adversarialattacks,
    author      = {Lea Sch{\"{o}}nherr and
                  Katharina Kohls and
                  Steffen Zeiler and
                  Thorsten Holz and
                  Dorothea Kolossa},
    title       = {Adversarial Attacks Against Automatic Speech Recognition Systems via Psychoacoustic Hiding},
    booktitle   = {Proc. of the 26th NDSS},
    year        = {2019},
    pages       = {1--18},
}

@inproceedings{Ghost13,
  author       = {Denis Foo Kune and
                  John D. Backes and
                  Shane S. Clark and
                  Daniel B. Kramer and
                  Matthew R. Reynolds and
                  Kevin Fu and
                  Yongdae Kim and
                  Wenyuan Xu},
  title        = {Ghost Talk: Mitigating {EMI} Signal Injection Attacks against Analog
                  Sensors},
  booktitle    = {{IEEE} Symposium on Security and Privacy},
  pages        = {145--159},
  year         = {2013}
}

@ARTICLE{IEMI15,
  author={Kasmi, Chaouki and Lopes Esteves, Jose},
  journal={IEEE Transactions on Electromagnetic Compatibility}, 
  title={IEMI Threats for Information Security: Remote Command Injection on Modern Smartphones}, 
  volume={57},
  number={6},
  pages={1752-1755},
  year={2015}

}

@inproceedings{GhostTalk22,
  author       = {Yuanda Wang and
                  Hanqing Guo and
                  Qiben Yan},
  title        = {GhostTalk: Interactive Attack on Smartphone Voice System Through Power
                  Line},
  booktitle    = {29th Annual Network and Distributed System Security Symposium},
  year         = {2022}
}

@inproceedings{DualGuard20,
  author       = {Shu Wang and
                  Jiahao Cao and
                  Xu He and
                  Kun Sun and
                  Qi Li},
  title        = {When the Differences in Frequency Domain are Compensated: Understanding
                  and Defeating Modulated Replay Attacks on Automatic Speech Recognition},
  booktitle    = {Proc. of the ACM SIGSAC
Conference on Computer and Communications Security},
  pages        = {1103--1119},
  year         = {2020}
}

@inproceedings{Cacher19,
  author       = {Chen Yan and
                  Yan Long and
                  Xiaoyu Ji and
                  Wenyuan Xu},
  title        = {The Catcher in the Field: {A} Fieldprint based Spoofing Detection
                  for Text-Independent Speaker Verification},
  booktitle    = {Proc. of the 26th ACM CCS},
  pages        = {1215--1229},
  year         = {2019}
}

@inproceedings{VAuth17,
  author       = {Huan Feng and
                  Kassem Fawaz and
                  Kang G. Shin},
  title        = {Continuous Authentication for Voice Assistants},
  booktitle    = {Proc. of ACM MobiCom },
  pages        = {343--355},
  publisher    = {{ACM}},
  year         = {2017}
}

@inproceedings{RobustDetection21,
  author       = {Zhuohang Li and
                  Cong Shi and
                  Tianfang Zhang and
                  Yi Xie and
                  Jian Liu and
                  Bo Yuan and
                  Yingying Chen},
  title        = {Robust Detection of Machine-induced Audio Attacks in Intelligent Audio
                  Systems with Microphone Array},
  booktitle    = {Proc. of the 28th ACM CCS},
  pages        = {1884--1899},
  year         = {2021}
}

@inproceedings{arrayID22,
  author       = {Yan Meng and
                  Jiachun Li and
                  Matthew Pillari and
                  Arjun Deopujari and
                  Liam Brennan and
                  Hafsah Shamsie and
                  Haojin Zhu and
                  Yuan Tian},
  title        = {Your Microphone Array Retains Your Identity: {A} Robust Voice Liveness
                  Detection System for Smart Speakers},
  booktitle    = {31st {USENIX} Security Symposium},
  pages        = {1077--1094},
  year         = {2022}
}

@article{survey21,
  author       = {Anirban Chakraborty and
                  Manaar Alam and
                  Vishal Dey and
                  Anupam Chattopadhyay and
                  Debdeep Mukhopadhyay},
  title        = {A survey on adversarial attacks and defences},
  journal      = {{CAAI} Trans. Intell. Technol.},
  volume       = {6},
  number       = {1},
  pages        = {25--45},
  year         = {2021}
}

@inproceedings{Brother20,
  author       = {Yue Huang and
                  Borke Obada{-}Obieh and
                  Konstantin Beznosov},
  title        = {Amazon vs. My Brother: How Users of Shared Smart Speakers Perceive
                  and Cope with Privacy Risks},
  booktitle    = {ACM CHI},
  pages        = {1--13},
  year         = {2020}
}

\begin{IEEEbiography}[{\includegraphics[width=1in, height=1.25in, clip, keepaspectratio]{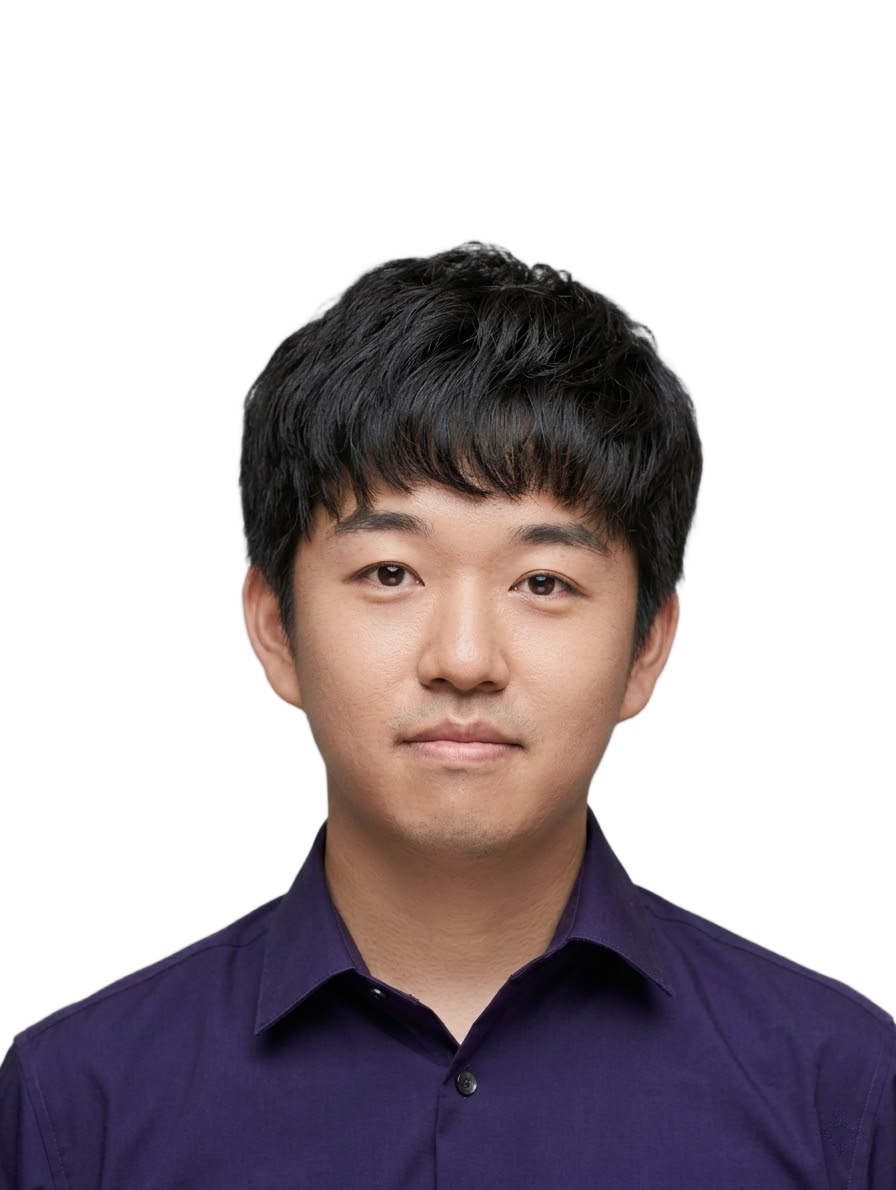}}]{Chao Liu} (Senior Member, IEEE) received the B.S. degree from the Ocean University of China, Qingdao, China, in 2011 and the Ph.D. degree from the Ocean University of China in 2016. He is currently a Full Professor with the School of Computer Science and Technology, Ocean University of China. His main research interests include acoustic sensing, mobile computing, and wireless sensing. He has authored or co-authored more than 100 papers in top journals and conference proceedings, such as the CCS, INFOCOM, UbiComp, ICDCS, IEEE JOURNAL ON SELECTED AREAS IN COMMUNICATIONS, IEEE TRANSACTIONS ON MOBILE COMPUTING, and IEEE TRANSACTIONS ON IMAGE PROCESSING. He was awarded the ACM CCS Distinguished Paper Award for proposing the RefleXnoop keystroke attack model.
\end{IEEEbiography}
\vfill
 \begin{IEEEbiography}[{\includegraphics[width=1in, height=1.25in, clip, keepaspectratio]{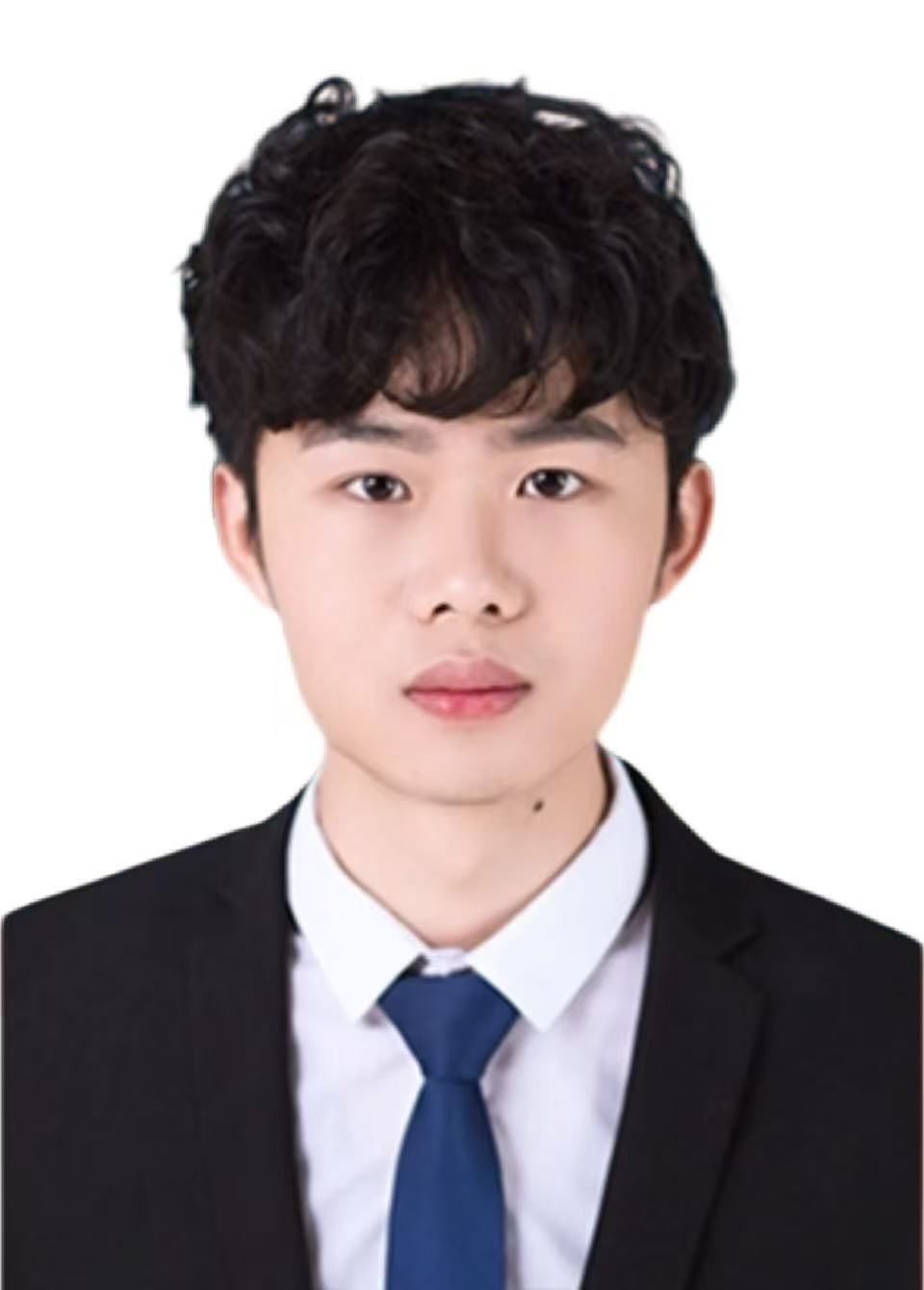}}]{Zhezheng Zhu} is pursuing the  master degree in the Department of Computer Science and Technology,  Ocean University of China,  Qingdao China. His main research interests include acoustic sensing. His main research interests include wireless sensing and wireless sensor networks.
 \end{IEEEbiography}

 \begin{IEEEbiography}[{\includegraphics[width=1in, height=1.25in, clip, keepaspectratio]{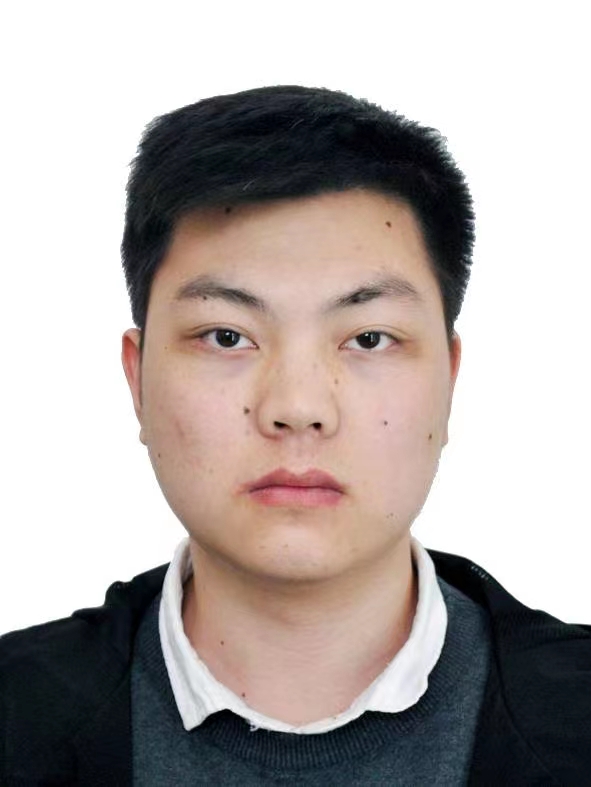}}]{Hao Chen} is pursuing the Ph.D. degree in the Department of Computer Science and Technology,  Ocean University of China,  Qingdao China. His main research interests include wireless sensing and wireless sensor networks. He has published papers in IEEE INFOCOM, IEEE ICDCS, and IEEE TMC. 
 \end{IEEEbiography}
  
\begin{IEEEbiography}[{\includegraphics[width=1in, height=1.25in, clip, keepaspectratio]{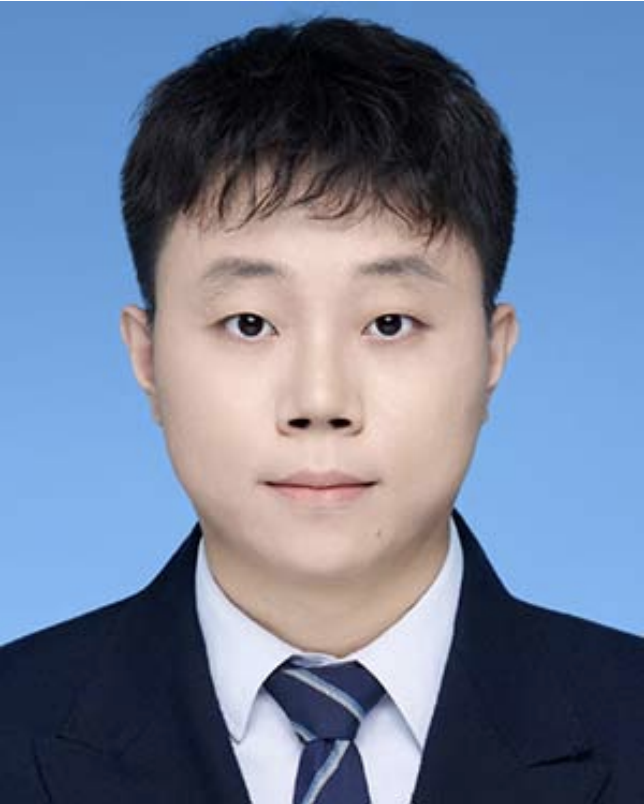}}]{Kaiwen Guo} received the BE degree from the Hefei University of Technology, Hefei, China, in 2019, and the PhD degree in computer science from the University of Science and Technology of China, Hefei, in 2024. He is currently an lecturer with the Ocean University of China, Qingdao, China. His research interests include wireless sensing and mobile computing.
 \end{IEEEbiography}
 
 \begin{IEEEbiography}[{\includegraphics[width=1in, height=1.25in,clip,keepaspectratio]{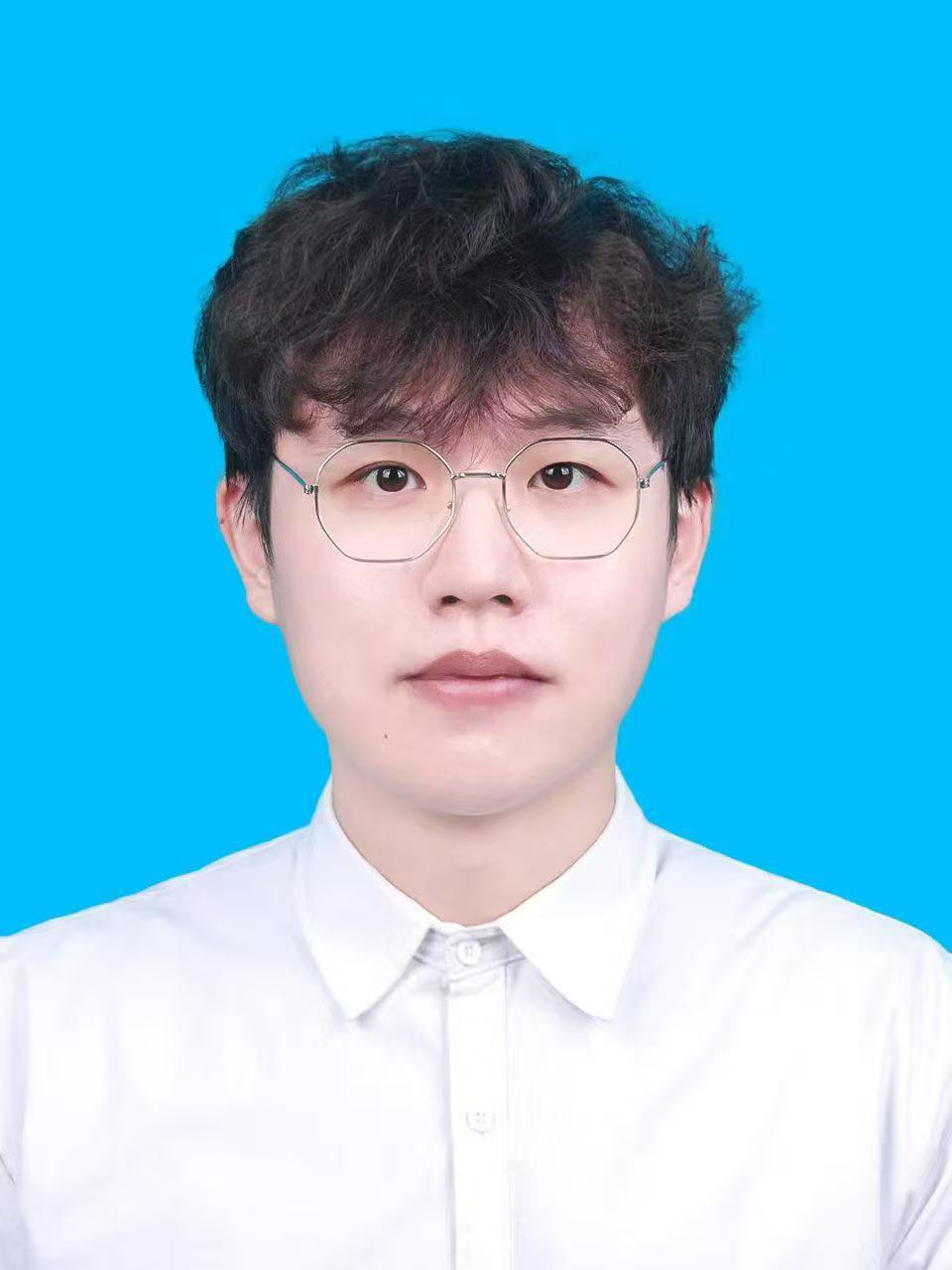}}]{Penghao Wang} received the Ph.D. degree in computer science and technology from Ocean University of China, Qingdao, China. He is currently a Research Fellow with Nanyang Technological University, Singapore. His research focuses on acoustic and ultrasonic sensing, wireless sensing, human–computer interaction, and sensing security. He has published several papers in top-tier venues, including ACM CCS, ACM UbiComp, IEEE INFOCOM, IEEE TMC, and ACM TOSN.
 \end{IEEEbiography}

 \begin{IEEEbiography}[{\includegraphics[width=1in, height=1.25in, clip, keepaspectratio]{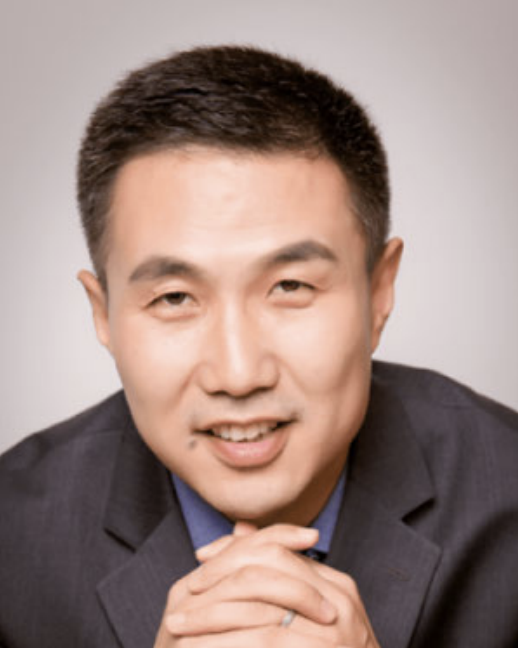}}]{Xiang-Yang Li} (Fellow, IEEE) received the bachelor’s degree from the Department of Computer Science, Tsinghua University, China, in 1995, and the M.S. and Ph.D. degrees from the Department of Computer Science, University of Illinois Urbana–Champaign, in 2000 and 2001, respectively. He was a Full Professor with the Computer Science Department, IIT. He is currently a Professor and the Executive Dean of the School of Computer Science and Technology, USTC. His research interests include the artificial intelligence of things (AIOT), privacy and security of AIOT, and data sharing and trading. He was an ACM Fellow in 2019 and an ACM Distinguished Scientist in 2014.
 \end{IEEEbiography}
\vfill
\end{document}